\documentclass [11pt]{article}
\usepackage[dvips]{graphicx}
\usepackage{subcaption}

\usepackage{epsfig}
\usepackage{graphicx}
\usepackage{amsmath}
\usepackage{amsfonts}
\usepackage{amssymb}
\usepackage{hyperref}
\usepackage{latexsym}
\usepackage{color}

\begin{document}
%%%%%%%%%%%%%%%%%%%%%%%%%%%%%%%%%%%%%%%%%%%%%%%%%%%%%%%%%%%%%%%%%%%%%%%%

\begin{center}
\begin{flushright}
	\begin{small}    
\end{small}
 \end{flushright} \vspace{1.5cm}
\Large{\bf Periodic cosmic evolution in Hybrid and Logarithmic Teleparallel Gravity}
\end{center}

\begin{center}
F. Mavoa $^{(a,b)}$\footnote{e-mail: ferdinand.mavoa@univ-labe.edu.gn, maferdson@yahoo.fr},
M.C . Sow $^{(c)}$\footnote{recteur@univ-labe.edu.gn
	}
H. Hova 
$^{(d)}$\footnote{hovhoav@gmail.com},
C. S. Tour\'e $^{(a,f)}$\footnote{e-mail: cheikhtoure.@univ-labe.edu.gn},
M. G. Ganiou $^{(e)}$\footnote{ganiou.gbenga@uganc.edu.gn},

\vskip 4mm
$^{a}$\, {\it  D\'epartement Energie Photovolta\"{i}que, Universit\'e de Lab\'e, R\'epublique de Guin\'ee}\\ 

{\it B.P:(+224) 2010,  Lab\'e, R\'epublique de Guin\'ee}\\

$^b$ \,{\it International Chair of Mathematical Physics and Application (ICMPA) , University d'Abomey-Calavi, B\'enin}\\
 {\it  072 BP 50, Cotonou, B\'enin}\\
 
 $^{c}$\, {\it  Facult\'e des Sciences Techniques (FST), Universit\'e de Lab\'e, R\'epublique de Guin\'ee}\\
 {\it B.P:(+224) 2010,  Lab\'e, R\'epublique de Guin\'ee}\\
 
$^d$ \,{\it D\'epartement de Physique, Facult\'e des Sciences et techniques, Universit\'e de Kara}\\
{\it BP:404 Kara ,  Togo}\\

$^{e}$\, {\it  D\'epartementde Phsique , Universit\'e Gamal Abdel Nasser, R\'epublique de Guin\'ee}\\
{\it BP 1147 Conakry , R\'epublique de Guin\'ee}\\

$^{f}$\, {\it  \'equipe de Recherche Energies Renouvelables Matériaux et Lasers , Universit\'e Alioune Diop, R\'epublique de S\'enegal}\\
{\it BP 30 Bambey , R\'epublique de S\'enegal}\\

\vskip 2mm
\end{center}

%%%%%%%%%%%%%%%%%%%%%%%%%%%%%%%%%%%%%%%%%%%%%%%%%%%%%%%%%%%%%%%%%%%%%%%%%%%%%%%%%%%%%%%%%%%%%%%%%%%%%%%%%%%%%%%%%%%%%%%%%%%%%%%%%%%%%%%%%%%%%%%%%%%%%%%

\begin{abstract}
	
In this work, we investigate a cosmological model within the framework of modified teleparallel gravity by considering two functional forms of $f(T)$: a hybrid model $f(T)=e^{\gamma T} T^{\sigma}$ and a logarithmic model, in the context of a periodic cosmic evolution driven by an oscillating deceleration parameter $q(t)=m\cos(kt)-1$. This parametrization allows for a cyclic description of the Universe, characterized by successive transitions between decelerating and accelerating phases.
		
By constraining the model with observational values such as $m \simeq 0.48$ and $H_0 = 69.2\,\text{km}\,\text{s}^{-1}\,\text{Mpc}^{-1}$, we show that it successfully reproduces the present accelerated expansion with $q_0 \approx -0.52$, while larger values $m \geq 1$ lead to strongly oscillatory regimes including super-acceleration phases.
		
For the hybrid model with $\gamma = 0.1$ and $\sigma = -0.5$, the energy density remains positive throughout cosmic evolution, whereas the pressure exhibits oscillatory behavior. The equation of state parameter dynamically evolves and crosses both quintessence and phantom regimes, highlighting the model's ability to describe different dark energy phases.
		
In contrast, the logarithmic model introduces a stabilizing effect on the cosmological dynamics. It regularizes divergences arising near singular points and leads to smoother evolutions of physical quantities. In particular, the equation of state parameter shows reduced oscillations and remains predominantly in the quintessence regime, thereby limiting extreme phantom behavior.
		
The analysis of energy conditions (NEC, DEC, SEC) reveals a strong dependence on model parameters. The violation of the SEC supports the accelerated expansion, while the partial validity of NEC and DEC ensures the physical consistency of the model during certain cosmological epochs.
		
Overall, this study demonstrates that the combination of periodic cosmic evolution with hybrid and logarithmic corrections in $f(T)$ gravity provides a rich and flexible framework, offering a viable alternative to the standard $\Lambda$CDM model and a unified description of the different phases of cosmic expansion.

\end{abstract}

%%%%%%%%%%%%%%%%%%%%%%%%%%%%%%%%%%%%%%%%%%%%%%%%%%%%%%%%%%%%%%%%%%%%%%%%%%%%%%%%%%%%%%%%%%%%%%%%%%%%%%%%%%%%%%%%%%%%%%%%%%%%%%%%%%%%%%%%%%%%%%%%%%%%%%%

%%%%%%%%%%%%%%%%%

%%%%%%%%%%%%%%%%%%%%%%%%%%%%%%%%%%%%%%%%%%%%%%%%%%%%%%%%%%%%%%%%%%%%%%%%%%%%%%%%%%%%%%%%%%%%%%%%%%%%%%%%%%%%%%%%%%%%%%%%%%%%%%%%%%%%%%%%%%%%%%%%%%%%%%%
%\tableofcontents

\section{Introduction}
Cosmology aims to explain the origin of the Universe, its dynamical evolution over time, and its ultimate fate. Fundamental questions related to the Big Bang, cosmic inflation, dark matter, and dark energy reflect the complexity and richness of cosmic history. According to current theoretical frameworks, the Universe originated approximately 13 billion years ago from an extremely hot and dense initial state commonly referred to as the Big Bang.

One of the most compelling observational evidences supporting this primordial phase is the Cosmic Microwave Background (CMB), discovered by Penzias and Wilson \cite{Penzias1965}. Detailed measurements of the temperature anisotropies of the CMB have played a crucial role in establishing and refining the standard cosmological model. This model describes a Universe predominantly composed of cold dark matter and dark energy, the latter often characterized by the cosmological constant $\Lambda$, which is responsible for the observed accelerated expansion of the Universe \cite{Riess1998,Perlmutter1999}.

Within this framework, the $\Lambda$CDM model \cite{Peebles2003} is currently regarded as the most successful and widely accepted description of the expanding Universe. Nevertheless, numerous alternative models of dark energy have been proposed in the literature in order to test and extend the standard paradigm. These alternatives provide viable approaches to distinguish various cosmological scenarios from the $\Lambda$CDM model.

In this context, the Statefinder diagnostic has proven to be a powerful tool for investigating the expansion dynamics of the Universe and the nature of dark energy \cite{Sahni2003}. Its effectiveness has been examined in relation to the proposed SNAP satellite mission, which was expected to observe approximately 2000 supernovae per year. The analysis demonstrates that the Statefinder parameters are sufficiently sensitive to discriminate between a wide range of dark energy models, including the cosmological constant, quintessence, Chaplygin gas, and braneworld scenarios.

Furthermore, the application of the Statefinder diagnostic has led to the reconstruction of holographic dark energy models \cite{Li2004,Gao2009}. In addition, modified theories of gravity, such as $f(R)$, $f(G)$, and $f(T)$ gravity, have been introduced as alternative explanations for cosmic acceleration, offering a geometrical interpretation of dark energy \cite{Nojiri2007,DeFelice2010,Cai2016}. These theoretical frameworks encompass a broad spectrum of cosmological models, including $\Lambda$CDM cosmology, Little Rip and Pseudo-Rip universes, as well as phantom and quintessence cosmologies, which may lead to various types of finite-time future singularities or to non-singular dark energy universes.

In the present work, we investigate the modified teleparallel theory of gravity $f(T)$, which constitutes an alternative extension of General Relativity based on spacetime torsion rather than curvature. In this framework, the gravitational interaction is described by the torsion scalar $T$, allowing the late–time cosmic acceleration to be interpreted as a purely geometric effect without explicitly introducing exotic dark energy components. Modified teleparallel models have been shown to provide viable cosmological scenarios capable of explaining the observed accelerated expansion \cite{Cai2016,Bengochea2009,Linder2010,Bamba2013}. In order to enrich the phenomenology and achieve greater dynamical flexibility, we adopt here a hybrid functional form of the gravitational Lagrangian that combines both polynomial and logarithmic contributions of $T$. Such a choice allows a broader class of cosmological behaviors and improves the description of the Universe at different evolutionary stages.

To describe the large–scale cosmological dynamics within this theoretical setting, we adopt the homogeneous and isotropic Friedmann–Lemaître–Robertson–Walker (FLRW) geometry. From a mathematical point of view, this spacetime can be interpreted as a warped product between the temporal dimension and a three–dimensional space of constant curvature. The global evolution of the Universe is fully determined by the scale factor $a(t)$, which encodes the history of cosmic expansion.

From this fundamental quantity, several kinematical parameters are introduced in order to characterize the expansion dynamics. In particular, the expansion rate is quantified through the Hubble parameter
\begin{equation}
H(t)=\frac{\dot a}{a},
\end{equation}
which measures the relative rate of expansion or contraction of the Universe.

The study of cosmic acceleration further involves the deceleration parameter
\begin{equation}
q=-\frac{a\,\ddot a}{\dot a^{2}},
\end{equation}
which distinguishes accelerating phases from decelerating ones.

Whenever $\ddot a>0$, the Universe undergoes accelerated expansion, corresponding to negative values of $q$. The minus sign in its definition has a historical origin: before the discovery of cosmic acceleration, the Universe was assumed to be decelerating ($\ddot a<0$), and this convention ensured positive values of $q$. However, current observations indicate $q_0<0$, confirming the present accelerated phase.

In general, the deceleration parameter evolves with cosmic time, although some simplified models assume it to be constant. For instance, a particular law for the variation of the Hubble parameter leading to a constant $q$ was proposed by Berman and collaborators \cite{Berman1983,BermanGomide1988}, which can serve as an approximation for slowly varying regimes. In the standard $\Lambda$CDM scenario, $q$ typically evolves between $1/2$ and $-1$. To account for such behavior, several parametrizations have been considered, including linear and quadratic dependences on cosmic time or redshift $z$, as well as more general Taylor–series expansions.

In the present analysis, we instead assume a periodic law for the deceleration parameter. This type of behavior was first explored in the context of General Relativity and later in modified gravity theories, including $f(R,T)$ gravity \cite{Shen2014} and more recently in the non–metricity based $f(Q)$ gravity framework \cite{Jimenez2018}\cite{Sahoo2022}
. Here, we extend this approach to the context of $f(T)$ gravity.

%%%%%%%%%%%%%%%%%%%%%%%%%%%%%%%%%%%%%%%%%%%%%%%%%%%%%%%%%%%%

\section{Main equations in the coupling-modified teleparallel theory with scalar field}

The modified $f(T)$ theory of gravity is based on a consistent and well-established mathematical
framework. In this section, we briefly summarize the fundamental aspects of teleparallel gravity
that are relevant to the present investigation.

In gravitational theories, the metric tensor plays a central role since it provides the necessary
information to measure spacetime intervals locally and to derive observable predictions.
Alternatively, the spacetime geometry can be described using a different set of dynamical variables,
namely the tetrad fields $h^{a}{}_{\mu}$, which form a set of four linearly independent vectors
defining a local orthonormal frame at each spacetime point. These tetrads represent the fundamental
variables of teleparallel gravity.

Teleparallel gravity can be interpreted as a gauge theory of the translation group. Within this
formalism, the tetrad field arises from the gauge covariant derivative acting on a scalar field and
is defined as
\begin{equation}
h^{a}{}_{\mu} = \partial_{\mu} x^{a} + A^{a}{}_{\mu},
\end{equation}
where $A^{a}{}_{\mu}$ denotes the translational gauge potential and $x^{a}$ are the tangent-space
coordinates \cite{AndradePereira1997}
.

The tetrad field $h^{a}{}_{\mu}$ and its inverse $h_{a}{}^{\mu}$ satisfy the orthonormality relations
\begin{equation}
h^{a}{}_{\mu} h_{a}{}^{\nu} = \delta^{\nu}_{\mu},
\qquad
h^{a}{}_{\mu} h_{b}{}^{\mu} = \delta^{a}_{b}.
\label{eq:tetrad_relations}
\end{equation}

An essential feature of teleparallel gravity is the condition of absolute parallelism
\cite{HayashiShirafuji1979}
, which leads to the introduction of the Weitzenb\"ock connection as the fundamental
connection of the theory. It is given by
\begin{equation}
\Gamma^{\lambda}{}_{\mu\nu}
= h_{a}{}^{\lambda} \partial_{\nu} h^{a}{}_{\mu}
= - h^{a}{}_{\mu} \partial_{\nu} h_{a}{}^{\lambda}.
\label{eq:weitzenbock}
\end{equation}

Throughout this work, Latin indices $(a,b,c,\ldots=0,1,2,3)$ label the tangent space, while Greek
indices $(\mu,\nu,\rho,\ldots=0,1,2,3)$ denote spacetime coordinates. The spacetime metric is related
to the tetrad fields by
\begin{equation}
g_{\mu\nu} = \eta_{ab} h^{a}{}_{\mu} h^{b}{}_{\nu},
\label{eq:metric_tetrad}
\end{equation}
where $\eta_{ab}=\mathrm{diag}(+1,-1,-1,-1)$ is the Minkowski metric of the tangent space.

In teleparallel gravity, the curvature associated with the Weitzenb\"ock connection vanishes
identically, and gravitational effects are entirely encoded in the torsion tensor, while the
curvature tensor does not appear. The torsion tensor is defined as
\begin{equation}
T^{\lambda}{}_{\mu\nu}
= \Gamma^{\lambda}{}_{\nu\mu} - \Gamma^{\lambda}{}_{\mu\nu}
= h_{a}{}^{\lambda}
\left(
\partial_{\mu} h^{a}{}_{\nu}
- \partial_{\nu} h^{a}{}_{\mu}
\right)
\neq 0.
\label{eq:torsion}
\end{equation}

Another important geometrical quantity is the contortion tensor
$K^{\lambda}{}_{\mu\nu}$, which measures the difference between the Weitzenb\"ock and Levi--Civita
connections \cite{Hehl1976,AldrovandiPereira2013}
. It is defined by
\begin{equation}
K^{\lambda}{}_{\mu\nu}
= \Gamma^{\lambda}{}_{\mu\nu}
- \tilde{\Gamma}^{\lambda}{}_{\mu\nu}
= \frac{1}{2}
\left(
T_{\nu}{}^{\lambda}{}_{\mu}
+ T_{\mu}{}^{\lambda}{}_{\nu}
- T^{\lambda}{}_{\mu\nu}
\right),
\label{eq:contortion}
\end{equation}
where $\tilde{\Gamma}^{\lambda}{}_{\mu\nu}$ denotes the Levi--Civita connection.

The action of $f(T)$ gravity is given by
\begin{equation}
S = \frac{1}{4\kappa^{2}} \int d^{4}x \, h \, f(T)
+ \int d^{4}x \, h \, \mathcal{L}_{m},
\label{eq:action}
\end{equation}
where $h = |\det(h^{a}{}_{\mu})| = \sqrt{-g}$,
$\kappa^{2} = 16\pi G/c^{4}$, and $\mathcal{L}_{m}$ represents the matter Lagrangian.

By varying the action with respect to the tetrad fields, one obtains the field equations
\begin{equation}
\frac{1}{h}\partial_{\mu}
\left(h S_{a}{}^{\mu\nu}\right) f_{T}
- h_{a}{}^{\lambda} T^{\rho}{}_{\mu\lambda}
S_{\rho}{}^{\mu\nu} f_{T}
+ S_{a}{}^{\mu\nu} \partial_{\mu}(T) f_{TT}
+ \frac{1}{4} h_{a}{}^{\nu} f(T)
= \frac{1}{4\kappa^{2}} T_{a}{}^{\nu},
\label{eq:field_eq}
\end{equation}
where $f_{T} = df/dT$, $f_{TT} = d^{2}f/dT^{2}$, and $T_{a}{}^{\nu}$ is the energy--momentum tensor.

We consider a homogeneous and isotropic universe described by the
Friedmann--Lema\^{\i}tre--Robertson--Walker metric
\begin{equation}
ds^{2} = dt^{2}
- a^{2}(t)\left(dx^{2}+dy^{2}+dz^{2}\right),
\label{eq:flrw}
\end{equation}
where $a(t)$ is the scale factor. For this metric, the torsion scalar takes the form
\begin{equation}
T = -6H^{2},
\label{eq:torsion_scalar}
\end{equation}
with $H=\dot{a}/a$ being the Hubble parameter.

Assuming that the cosmic content is described by a perfect fluid driven by a scalar field $\phi$,
the energy--momentum tensor is written as
\begin{equation}
T_{\mu\nu}
= (\rho + p) u_{\mu} u_{\nu}
- p g_{\mu\nu},
\label{eq:emtensor}
\end{equation}
where $u_{\mu}$ is the four-velocity of a comoving observer, and $\rho$ and $p$ denote the energy
density and pressure, respectively.

Under these assumptions, the modified Friedmann equations in $f(T)$ gravity are obtained as
\begin{equation}
\kappa^{2} \rho
= 6H^{2} f_{T}
+ \frac{1}{4} f,
\label{eq:friedmann1}
\end{equation}
\begin{equation}
\kappa^{2} p
= 48 \dot{H} H^{2} f_{TT}
- (2\dot{H} + 6H^{2}) f_{T}
- \frac{1}{4} f.
\label{eq:friedmann2}
\end{equation}
	
\section{Cyclically evolving deceleration parameter}

The deceleration parameter \( q \) plays a crucial role in characterizing the dynamical evolution of the Universe, allowing one to distinguish between the past decelerated phase and the present accelerated expansion. The cosmic transition between these two regimes can be inferred from the sign change of the deceleration parameter.

In this work, we adopt a phenomenological periodic form of the deceleration parameter given by (\cite{Sahoo2022})
\begin{equation}
q(t) = m \cos(k t) - 1,
\label{eq:qparam}
\end{equation}
where $ m > 0 $ is a dimensionless constant that controls the amplitude of the oscillations, while $ k > 0 $ determines the periodicity of the cosmic evolution.

This parametrization allows for a rich dynamical behavior of the Universe. At early times, the Universe may experience a decelerated expansion phase characterized by a positive deceleration parameter. At late times, the parameter $ q $ can become negative, indicating an accelerated expansion, and may even reach a super-exponential regime when $ q < -1 $.

According to current cosmological observations, the present-day values of the deceleration parameter and the Hubble parameter are given by
\begin{equation}
q_0 = -0.52, \qquad H_0 = 69.2 \; \text{km s}^{-1}\text{Mpc}^{-1}.
\end{equation}

Imposing these observational constraints at the present cosmic time, the deceleration parameter (\ref{eq:qparam}) leads to a relation between the model parameters $ m $ and $ k $, expressed as
\begin{equation}
k = H_0 \cos^{-1}\!\left( \frac{q_0 + 1}{m} \right).
\label{eq:krelation}
\end{equation}

This relation restricts the allowed parameter space of the model and ensures consistency with late-time cosmic acceleration. Consequently, the periodic form of the deceleration parameter provides a viable phenomenological framework for exploring alternative cosmological scenarios and the dynamics of the accelerating Universe.

In the present model, the Universe initially undergoes a decelerating phase before transitioning into a phase of super-exponential expansion within a cyclic framework. This cyclic transitional behavior is relevant for values of $m \geq 1$. For $0 < m < 1$, the model preserves an accelerating phase and yields a present-day deceleration parameter around $m \approx 0.48$. The associated Hubble parameter can be expressed as:
	
	\begin{equation}
	H = \frac{k}{m \sin(k t) + \lambda},
	\end{equation}
	
	where $\lambda$ is an integration constant. Singularities may arise when the denominator vanishes for $|\lambda| \leq m$. Likewise, the Universe would undergo contraction for $\lambda < -m$; to avoid such issues, one should consider $\lambda > m$. For simplicity, and without loss of generality, we set $\lambda = 0$, yielding:
	
	\begin{equation}
	H = \frac{k}{m \sin(k t)}
	\label{eq:Hubble}
	\end{equation}

	Under the PVDP ansatz, the cosmic scale factor is obtained by integrating the Hubble function as:
	
	\begin{equation}
	a(t) = a_0 \left( \tan\frac{1}{2} k t \right)^{1/m},
	\end{equation}
	
	where $a_0$ represents the scale factor at the current epoch and can be normalized to $1$. Inverting this relation, the cosmic time as a function of redshift is:
	
	\begin{equation}
	t = \frac{2}{k} \tan^{-1} \left[ (1+z)^{-m} \right].
	\end{equation}
	
	These choices of $m$ and $k$ are subsequently used to explore the dynamical behavior of the model through numerical simulations.
\section{Solution}	
\subsection{Hybrid  Teleparallel Gravity}
For the first scenario, we assume a particular functional form of teleparallel gravity given by \cite{Mandal2020}

\begin{equation}
f(T) = e^{\gamma T} T^{\sigma},
\label{eq:fT_model}
\end{equation}
where $\gamma \geq 0 $ and $ \sigma $ are constant parameters. This model is especially interesting since it unifies both power-law and exponential behaviors depending on the specific choices of $\gamma $ and $ \sigma $. In particular:

\begin{itemize}
	\item For $ \gamma = 0 $, Eq.~(\ref{eq:fT_model}) reduces to the power-law form
	\begin{equation}
	f(T) = T^{\sigma}.
	\end{equation}
	
	\item For $ \sigma = 0 $, Eq.~(\ref{eq:fT_model}) simplifies to the exponential form
	\begin{equation}
	f(T) = e^{\gamma T}.
	\end{equation}
\end{itemize}

For the teleparallel gravity model defined by (\ref{eq:fT_model}) where $\gamma\ge $ and $\sigma$ are constants, the first and second derivatives of $f(T)$ with respect to the torsion scalar $T$ are given by
\begin{equation}
f_{T} = e^{\gamma T} T^{\sigma-1} (\sigma + \gamma T),
\label{eq:fT_first}
\end{equation}
and
\begin{equation}
f_{TT} = e^{\gamma T} T^{\sigma-2}
\left[n(n-1) + 2\gamma nT + \gamma^{2}T^{2}\right].
\label{eq:fT_second}
\end{equation}

Using the modified Friedmann equations of $f(T)$ gravity, (\ref{eq:friedmann1}) and (\ref{eq:friedmann2}) , the corresponding energy density and pressure for the present model take the explicit forms

\begin{equation}
\rho = \frac{e^{\gamma T} T^{\sigma-1}}{\kappa^{2}}
\left[
6H^{2}(\sigma + \gamma T) + \frac{1}{4}T
\right].
\label{eq:rho_factorized}
\end{equation}

and

\begin{equation}
\begin{aligned}
p = \frac{e^{\gamma T} T^{n-2}}{\kappa^{2}}
\Big[&
48 \dot{H} H^{2}
\left(\sigma(\sigma-1) + 2\gamma \sigma T + \gamma^{2}T^{2}\right) \\
&- (2\dot{H} + 6H^{2}) T (\sigma + \gamma T)
- \frac{1}{4} T^{2}
\Big].
\end{aligned}
\label{eq:p_factorized}
\end{equation}

Finally, the equation of state parameter is defined as
\begin{equation}
w = \frac{p}{\rho} =
\frac{ 
	48 \dot{H} H^{2} (\sigma(\sigma-1) + 2\gamma \sigma T + \gamma^{2}T^{2})
	- (2\dot{H} + 6H^{2}) T (\sigma + \gamma T)
	- \frac{1}{4} T^{2} 
}{
	T \left[ 6H^{2}(\sigma + \gamma T) + \frac{1}{4} T \right]
}.
\label{eq:w_factorized}
\end{equation}

We define the following parameters to simplify the notation:
	\begin{equation}
	\alpha = - \frac{6 k^2}{m^2}, \quad 
	\beta = \frac{k^2}{m^2}, \quad 
%	s = \sin(kt), \quad 
%	c = \cos(kt),
	\end{equation}

Using (\ref{eq:torsion_scalar}), (\ref{eq:Hubble}) the modified Friedmann equations, the energy density (\ref{eq:rho_factorized}) and pressure (\ref{eq:p_factorized}) and (\ref{eq:w_factorized})  can be written entirely in terms of $s$ and $c$:

\begin{equation}
\rho(t) = \frac{ e^{\gamma \alpha / \sin^2(kt)} \left( \frac{\alpha}{\sin^2(kt)} \right)^{\sigma -1} }{\kappa^2} 
\left[ 6 \frac{k^2}{m^2 \sin^2(kt)} \left( \sigma + \frac{\gamma \alpha}{\sin^2(kt)} \right)
+ \frac{1}{4} \frac{\alpha}{\sin^2(kt)} \right],
\label{eq:rho_t}
\end{equation}

\begin{equation}
\begin{aligned}
p(t) = \frac{ e^{\gamma \alpha / \sin^2(kt)} \left( \frac{\alpha}{\sin^2(kt)} \right)^{\sigma-2} }{\kappa^2} \Bigg[ &
48 \left( - \frac{k^2 \cos(kt)}{m \sin^2(kt)} \right)
\frac{k^2}{m^2 \sin^2(kt)} 
\left( \sigma(\sigma-1) + \frac{2 \gamma \sigma \alpha}{\sin^2(kt)} + \frac{\gamma^2 \alpha^2}{\sin^4(kt)} \right) \\
&- \left( -2 \frac{k^2 \cos(kt)}{m \sin^2(kt)} + 6 \frac{k^2}{m^2 \sin^2(kt)} \right) 
\left( \sigma + \frac{\gamma \alpha}{\sin^2(kt)} \right) \frac{\alpha}{\sin^2(kt)} \\
&- \frac{1}{4} \frac{\alpha^2}{\sin^4(kt)} \Bigg],
\end{aligned}
\label{eq:p_t}
\end{equation}

\begin{equation}
w(t) = \frac{
	\frac{1}{\alpha / \sin^2(kt)} 
	\Bigg[
	48 \frac{k^4}{m^3} \frac{-\cos(kt)}{\sin^6(kt)} 
	\left( \sigma(\sigma -1) + \frac{2 \gamma \sigma \alpha}{\sin^2(kt)} + \frac{m^2 \alpha^2}{\sin^4(kt)} \right)
	- \frac{k^2}{m^2} \frac{-2 \cos(kt) + 6/m}{\sin^4(kt)} 
	\left( \sigma + \frac{\gamma \alpha}{\sin^2(kt)} \right) \alpha
	- \frac{\alpha^2}{4 \sin^4(kt)}
	\Bigg]
}{
	6 \frac{k^2}{m^2 \sin^2(kt)} \left( \sigma + \frac{\gamma \alpha}{\sin^2(kt)} \right)
	+ \frac{\alpha}{4 \sin^2(kt)}
},
\label{eq:w_t}
\end{equation}
\clearpage
	\begin{figure}[h!]
		\centering
		
		\begin{subfigure}[b]{0.3\textwidth}
			\centering
			\includegraphics[width=\textwidth]{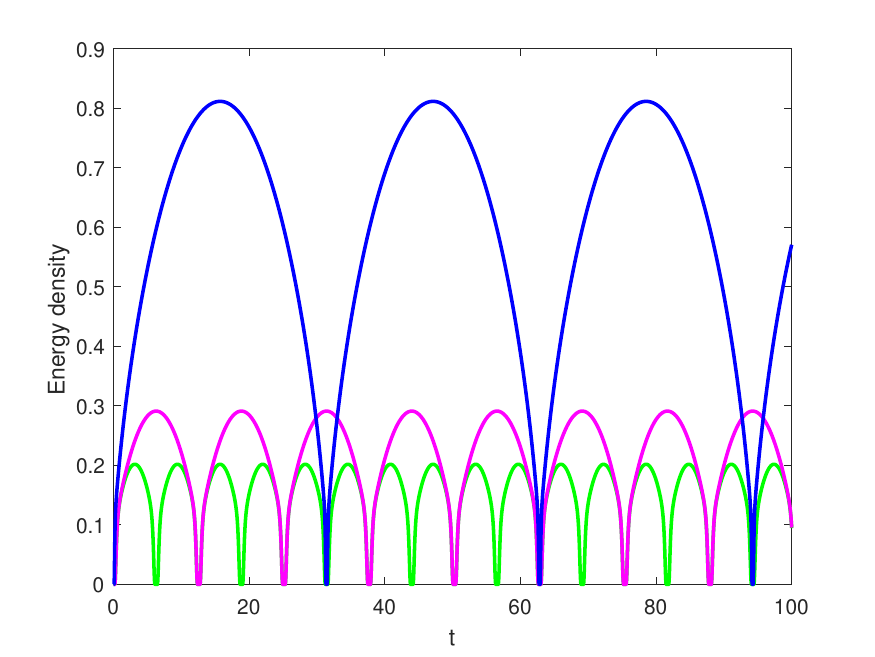}
			\caption{}
		\end{subfigure}
		\hfill
		\begin{subfigure}[b]{0.3\textwidth}
			\centering
			\includegraphics[width=\textwidth]{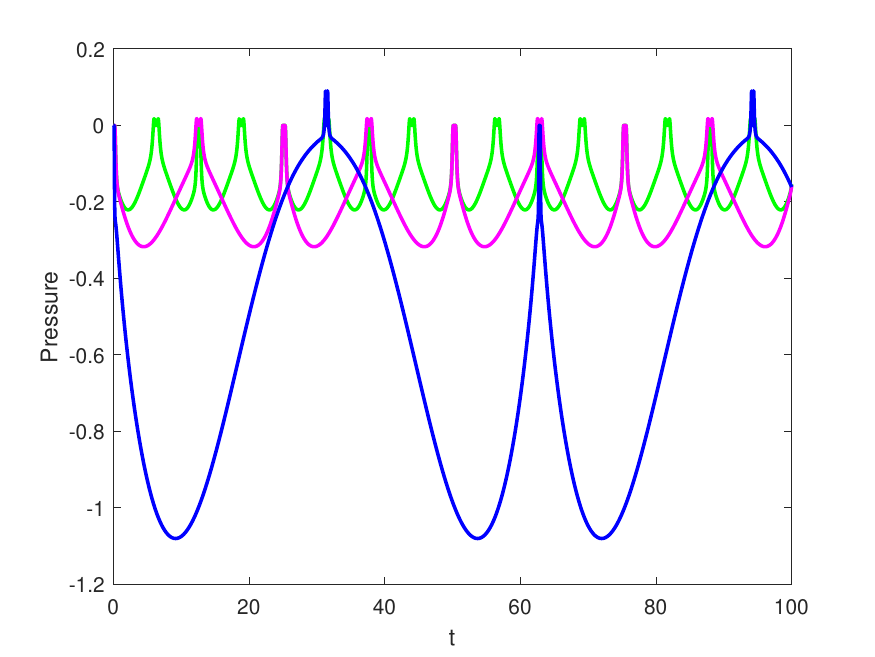}
			\caption{}
		\end{subfigure}
		\hfill
		\begin{subfigure}[b]{0.3\textwidth}
			\centering
			\includegraphics[width=\textwidth]{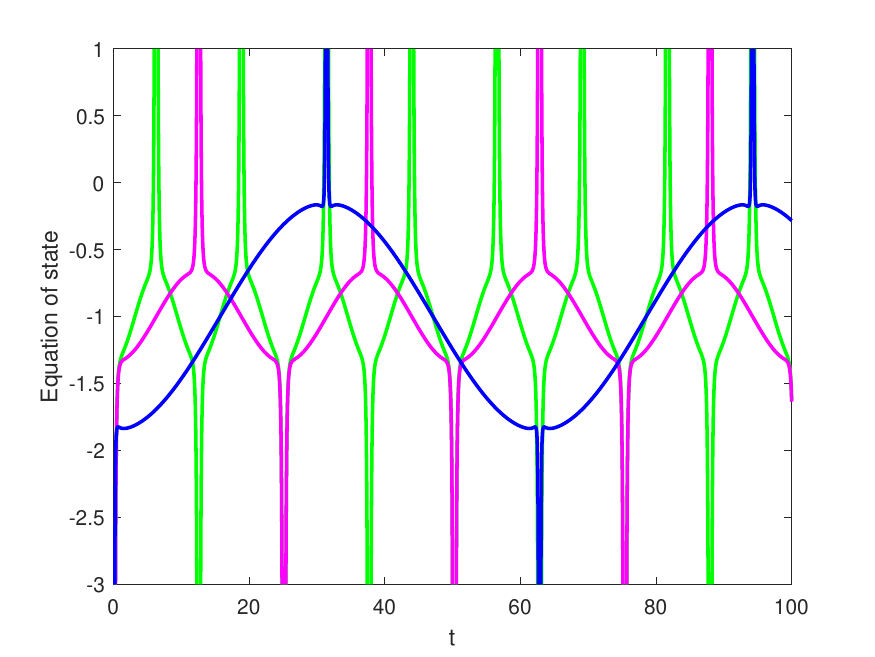}
			\caption{}
		\end{subfigure}
		
		\caption{Evolution of $\rho$, $p$, and $\omega$ as functions of cosmic time (in Gyr) for $\gamma = 0.1$, $\sigma = -0.5$. The results are shown for $(m, k) = (0.480012, 0.5)$ [green curve], $(m, k) = (0.480003, 0.25)$ [magenta curve], and $(m, k) = (1.15, 0.1)$ [blue curve].}
	\end{figure}
	
\begin{figure}[h!]
	\centering	
	\begin{subfigure}[b]{0.3\textwidth}
		\centering
		\includegraphics[width=\textwidth]{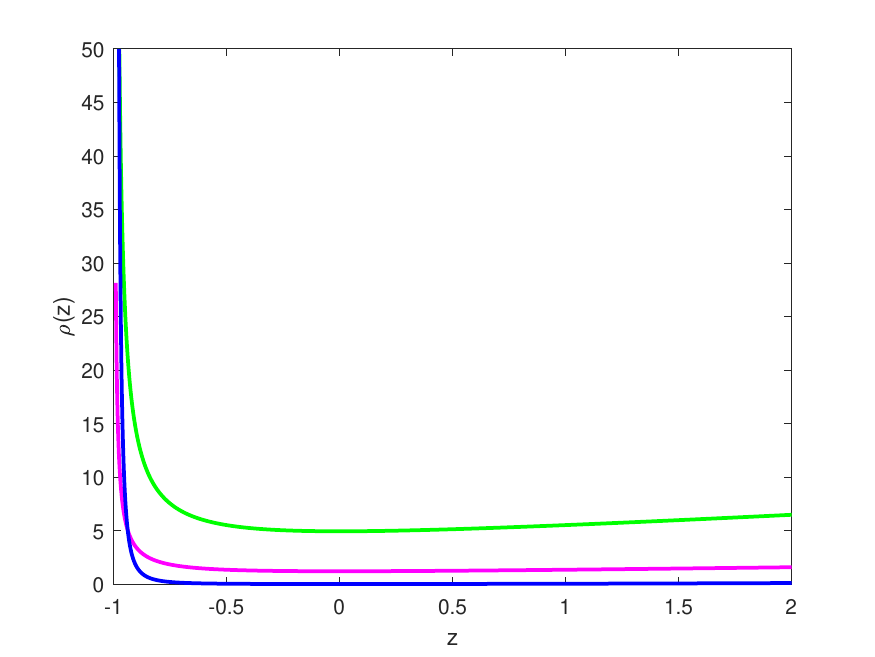}
		\caption{}
	\end{subfigure}
	\hfill
	\begin{subfigure}[b]{0.3\textwidth}
		\centering
		\includegraphics[width=\textwidth]{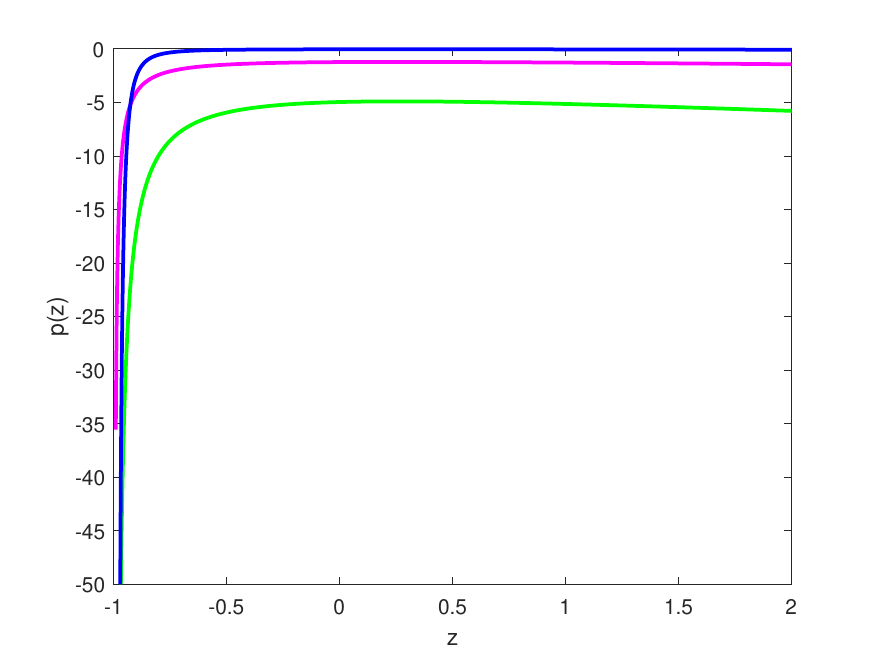}
		\caption{}
	\end{subfigure}
	\hfill
	\begin{subfigure}[b]{0.3\textwidth}
		\centering
		\includegraphics[width=\textwidth]{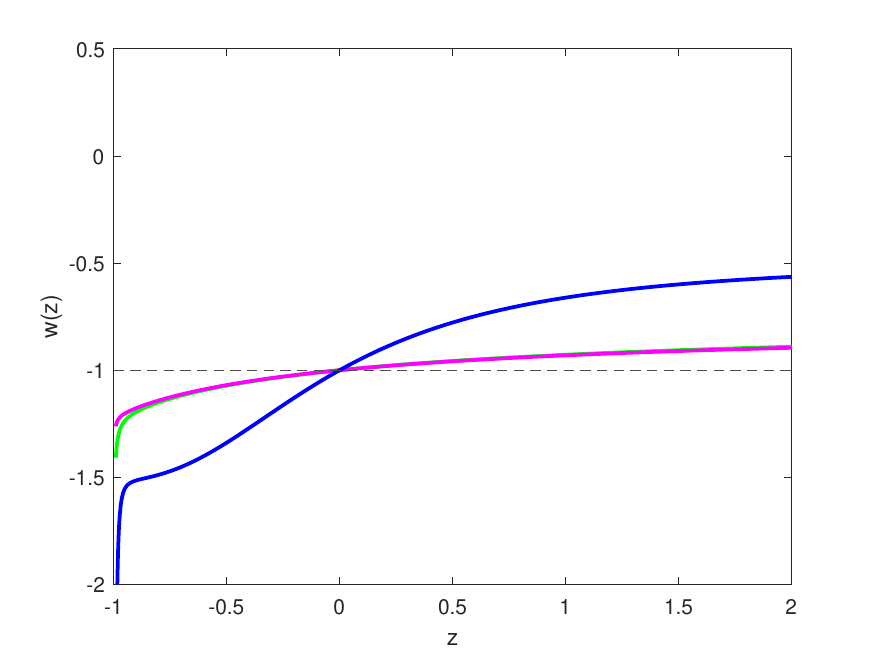}
		\caption{}
	\end{subfigure}
	
	\caption{Evolution of $\rho$, $p$, and $\omega$ as functions of redshift for $\gamma = 0.1$, $\sigma = -0.5$. The results correspond to $(m, k) = (0.480012, 0.5)$ [green curve], $(m, k) = (0.480003, 0.25)$ [magenta curve], and $(m, k) = (1.15, 0.1)$ [blue curve].}
\end{figure}
	
%\clearpage

	We first analyze the cosmological evolution within the framework of Hybrid Teleparallel Gravity for different choices of the parameter pairs $(m,k)=(0.480012,0.5)$ (green curve), $(0.480003,0.25)$ (magenta curve), and $(1.15,0.1)$ (blue curve).
	
	The evolution of the energy density $\rho(t)$ exhibits a clear periodic oscillatory behavior for all parameter choices. Notably, the energy density remains strictly positive throughout the evolution, ensuring the physical viability of the model. The amplitude and frequency of oscillations are strongly influenced by the parameters $(m,k)$. In particular, the blue curve corresponding to $(1.15,0.1)$ shows larger amplitude oscillations, indicating a more pronounced dynamical evolution, while the green and magenta curves display faster oscillations due to higher values of $k$. This behavior suggests a cyclic cosmological evolution characterized by repeated phases of expansion.
	
	The pressure $p(t)$ remains predominantly negative over the entire cosmic time for all considered parameter sets. This negative pressure is a clear signature of an accelerated expansion phase of the Universe. The oscillatory nature of the pressure indicates periodic transitions in the cosmic dynamics. The deeper negative values observed for the blue curve suggest a stronger effective dark energy component compared to the other cases.
	
	The equation of state (EoS) parameter $\omega(t)=p/\rho$ further confirms the accelerated behavior. The EoS parameter evolves periodically within the negative regime, oscillating between quintessence-like ($-1<\omega<0$) and phantom-like ($\omega<-1$) phases. The presence of sharp peaks or divergences in $\omega(t)$ indicates the occurrence of finite-time singularities, typically associated with Big Rip-type behavior. These singularities appear periodically, reflecting a cyclic Universe undergoing successive phases of expansion and extreme events.
	
	Overall, the Hybrid Teleparallel Gravity model describes a cosmological scenario in which the Universe evolves cyclically under the influence of a dominant negative pressure component. While the model successfully reproduces accelerated expansion and periodic behavior, the appearance of finite-time singularities suggests that additional physical mechanisms may be required to ensure a completely regular cosmic evolution.

\subsection{Logarithmic Teleparallel Gravity}
In the second scenario, we assume a logarithmic functional form for the teleparallel gravity theory, given by \cite{Mandal2020}

\begin{equation}
f(T) = D \ln(bT),
\label{eq:log_fT}
\end{equation}
where $D$ is a constant parameter and $b<0$.

We consider a logarithmic extension of teleparallel gravity (\ref{eq:log_fT}) for a spatially flat FRW universe, the torsion scalar is given by $T=-6H^2<0$.

The first and second derivatives of $f(T)$ with respect to $T$ read
\begin{equation}
f_T = \frac{D}{T}, \qquad
f_{TT} = -\frac{D}{T^2}.
\end{equation}
Substituting Eqs.~(\ref{eq:log_fT}) and the corresponding derivatives into Eq.~(\ref{eq:friedmann1})  the effective energy density and  becomes
\begin{equation}
\boxed{
	\rho = \frac{D}{\kappa^2}
	\left(
	\frac{1}{4}\ln(bT) - 1
	\right).
}
\label{eq:rho_log}
\end{equation}
Similarly, inserting the logarithmic form of $f(T)$ into Eq.~(\ref{eq:friedmann2}), we obtain the effective pressure
\begin{equation}
\boxed{
	p = \frac{D}{\kappa^2}
	\left(
	1 - \frac{\dot{H}}{H^2}
	- \frac{1}{4}\ln(bT)
	\right).
}
\label{eq:p_log}
\end{equation}

The equation-of-state parameter associated with the logarithmic $f(T)$ model is therefore given by
\begin{equation}
\boxed{
	w \equiv \frac{p}{\rho}
	=
	\frac{
		1 - \dfrac{\dot{H}}{H^2}
		- \dfrac{1}{4}\ln(bT)
	}{
		\dfrac{1}{4}\ln(bT) - 1
	}.
}
\label{eq:w_log}
\end{equation}

In this subsection, we derive the explicit time-dependent expressions of the effective
energy density and pressure and state equation in logarithmic teleparallel gravity.  From Eqs.~(35)-(37), we obtain the explicit expressions for the
energy density $\rho(t)$ [Eq.~(\ref{eq:rho_log})],
the pressure $p(t)$ [Eq.~(\ref{eq:p_log})], and the corresponding equation-of-state parameter $w(t)$ [Eq.~(\ref{eq:w_log})].

\begin{equation}
\boxed{
	\rho(t)
	=
	\frac{D}{4\kappa^2}
	\left[\ln\!\left(
	\frac{6|b|k^2}{m^2\sin^2(kt)}
	\right)
	-4
	\right].
}
\label{eq:rho_time}
\end{equation}

\begin{equation}
\boxed{
	p(t)
	=
	\frac{D}{4\kappa^2}
	\left[
	4
	+ 4 m\cos(kt)
	-
\ln\!\left(
	\frac{6|b|k^2}{m^2\sin^2(kt)}
	\right)
	\right].
}
\label{eq:p_time}
\end{equation}

\begin{equation}
\boxed{
	w(t)
	=
	\frac{
		4
		+ 4m\cos(kt)
		-
		\ln\!\left(
		\frac{6|b|k^2}{m^2\sin^2(kt)}
		\right)
	}{
		\ln\!\left(
		\frac{6|b|k^2}{m^2\sin^2(kt)}
		\right)
		-4
}.
}
\label{eq:w_time_simplified}
\end{equation}
In what follows, we present the cosmological evolution in the framework of Logarithmic Teleparallel Gravity through graphical illustrations, where we plot the behavior of the energy density $\rho$, the pressure $p$, and the equation of state parameter $w$ as functions of cosmic time as well as of the redshift $z$, in order to highlight the physical features of the model and assess its consistency with observational data.
\clearpage
\begin{figure}[h!]
	\centering
	
	\begin{subfigure}[b]{0.3\textwidth}
		\centering
		\includegraphics[width=\textwidth]{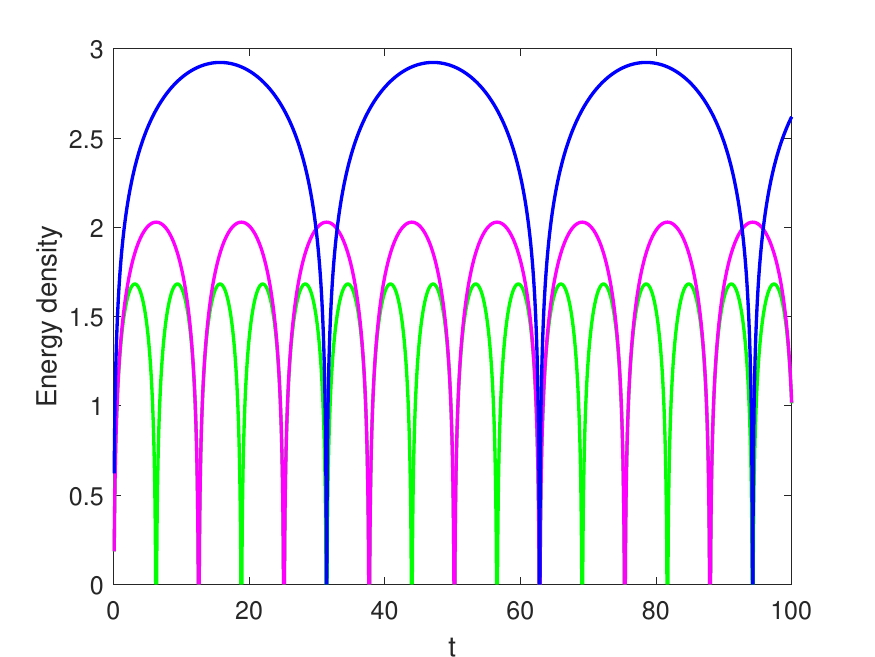}
		\caption{}
	\end{subfigure}
	\hfill
	\begin{subfigure}[b]{0.3\textwidth}
		\centering
		\includegraphics[width=\textwidth]{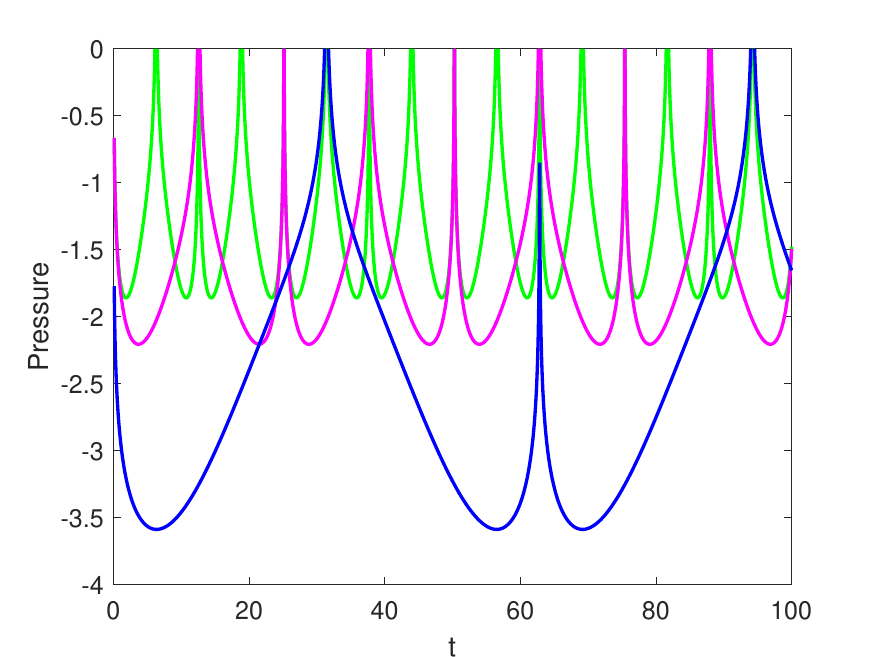}
		\caption{}
	\end{subfigure}
	\hfill
	\begin{subfigure}[b]{0.3\textwidth}
		\centering
		\includegraphics[width=\textwidth]{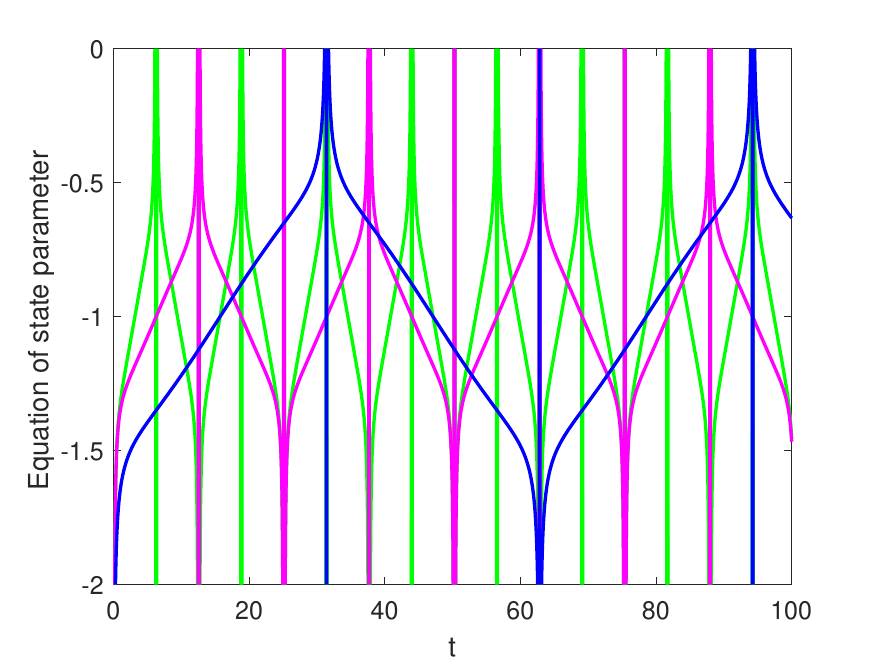}
		\caption{}
	\end{subfigure}
	
	\caption{Evolution of $\rho$, $p$, and $\omega$ as functions of cosmic time (in Gyr) for $D = 10$, $b = 1.5$. The results are shown for $(m, k) = (0.480012, 0.5)$ [green curve], $(m, k) = (0.480003, 0.25)$ [magenta curve], and $(m, k) = (1.15, 0.1)$ [blue curve].}
\end{figure}

\begin{figure}[h!]
	\centering	
	\begin{subfigure}[b]{0.3\textwidth}
		\centering
		\includegraphics[width=\textwidth]{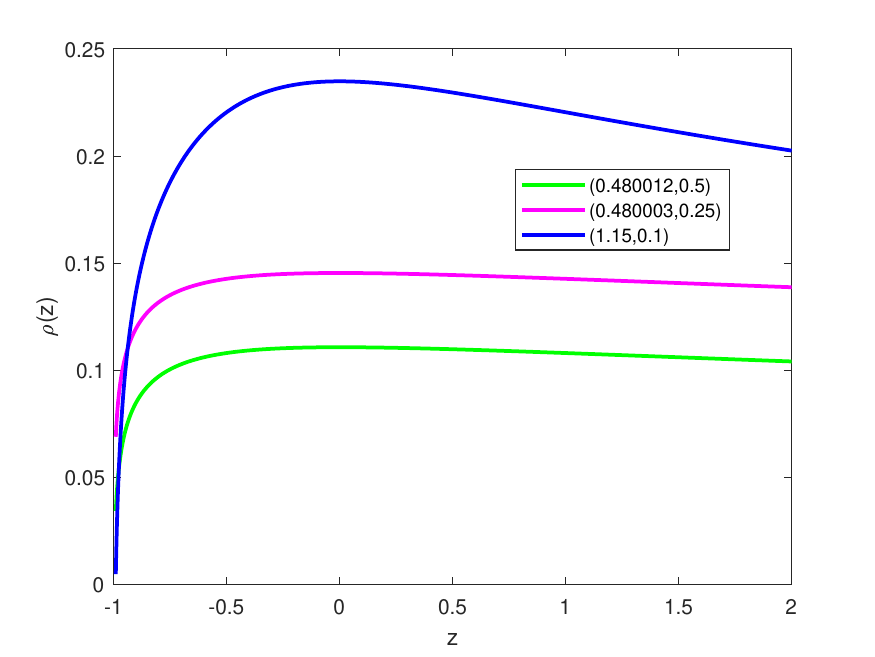}
		\caption{}
	\end{subfigure}
	\hfill
	\begin{subfigure}[b]{0.3\textwidth}
		\centering
		\includegraphics[width=\textwidth]{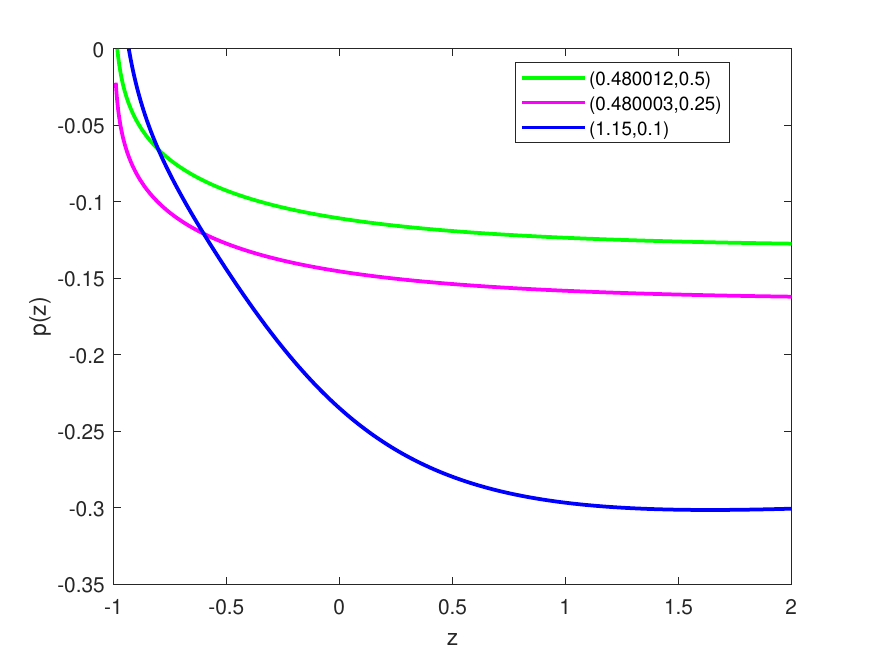}
		\caption{}
	\end{subfigure}
	\hfill
	\begin{subfigure}[b]{0.3\textwidth}
		\centering
		\includegraphics[width=\textwidth]{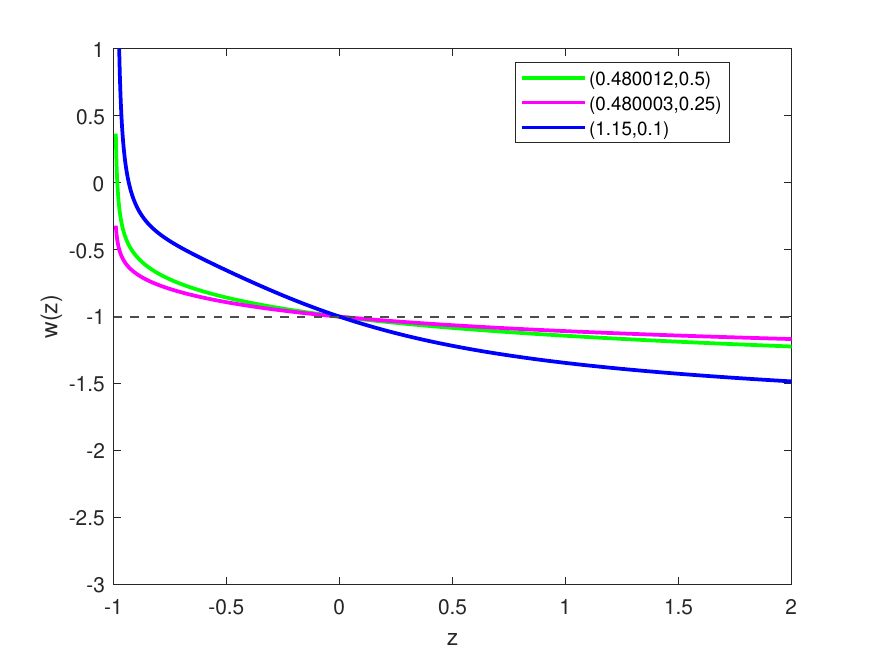}
		\caption{}
	\end{subfigure}
	
	\caption{Evolution of $\rho$, $p$, and $\omega$ as functions of redshift for $D = 10$, $b = 1.5$. The results correspond to $(m, k) = (0.480012, 0.5)$ [green curve], $(m, k) = (0.480003, 0.25)$ [magenta curve], and $(m, k) = (1.15, 0.1)$ [blue curve].}
\end{figure}

We now investigate the cosmological evolution in the framework of Logarithmic Teleparallel Gravity using the same parameter sets $(m,k)=(0.480012,0.5)$ (green curve), $(0.480003,0.25)$ (magenta curve), and $(1.15,0.1)$ (blue curve).

The energy density $\rho(t)$ exhibits a periodic behavior similar to the hybrid case, but with significantly enhanced amplitude. The density remains strictly positive for all parameter choices, confirming the physical consistency of the model. The logarithmic contribution in the gravitational action amplifies the oscillatory behavior, leading to stronger variations in the energy density. The blue curve again shows the largest amplitude, indicating a more energetic cosmological evolution, while the green and magenta curves correspond to higher frequency oscillations.

The pressure $p(t)$ is found to be negative throughout the evolution, indicating a sustained accelerated expansion phase. Compared to the hybrid case, the oscillations in pressure are more structured and regular. The stronger negative excursions observed, particularly for the blue curve, suggest a more dominant and persistent dark energy component.

The EoS parameter $\omega(t)$ displays periodic oscillations within a wider range, typically spanning from $\omega \approx -2$ to $\omega \approx -0.5$. This clearly indicates transitions between phantom and quintessence regimes during the cosmic evolution. As in the hybrid model, sharp divergences are observed in $\omega(t)$, corresponding to finite-time singularities of the Big Rip type. However, in this case, the singularities appear more regularly spaced, reinforcing the interpretation of a cyclic cosmological evolution.

An important feature of the logarithmic model is the enhanced regularity and symmetry of the oscillatory behavior. Despite the presence of singularities, the overall evolution appears more organized and dynamically richer than in the hybrid case. This suggests that logarithmic corrections in teleparallel gravity can significantly influence the structure of cosmic evolution, leading to stronger oscillations and more pronounced transitions between different dark energy regimes.

In conclusion, the Logarithmic Teleparallel Gravity model provides a more dynamically active cosmological scenario, characterized by amplified oscillations, persistent accelerated expansion, and periodic singularities, making it a compelling framework for describing cyclic cosmological behavior.

%\clearpage
\section{Energy Conditions in Hybrid and Logarithmic Teleparallel Gravity}

It is well established that the energy conditions play a fundamental role in characterizing the
attractive nature of gravity, while simultaneously determining the causal structure and the
geodesic behavior of spacetime \cite{Hawking1973,Visser1997}. Within the framework of generalized
teleparallel gravity $f(T)$, the analysis of energy conditions provides an important criterion
for testing the physical viability and cosmological consistency of the proposed models
\cite{Bamba2013,Cai2016}. By imposing these conditions, one can derive meaningful constraints
on the free parameters of the $f(T)$ function, ensuring realistic cosmic evolution.

The various energy conditions are widely employed in the study of cosmological dynamics.
In particular, the strong energy condition (SEC), defined by $\rho + 3p \geq 0$, together with
the weak energy condition (WEC), given by $\rho \geq 0$ and $\rho + p \geq 0$, have been used in the
formulation of the Hawking--Penrose singularity theorems \cite{Hawking1970}. Moreover, the null
energy condition (NEC), $\rho + p \geq 0$, is required to establish the validity of the second
law of black hole thermodynamics \cite{Bardeen1973}.

Originally formulated in the context of General Relativity, the energy conditions have been
extended to modified theories of gravity, including teleparallel gravity, by introducing
effective energy density and pressure terms arising from torsional contributions
\cite{Santos2016}. Furthermore, the dominant energy condition (DEC), expressed as
$\rho - p \geq 0$, is commonly regarded as a stability criterion, since it ensures that the
energy flux does not propagate faster than the speed of light, thereby preserving causality
\cite{Wald1984}.

In the present work, we investigate a hybrid and logarithmic formulation of teleparallel gravity,
in which the function $f(T)$ is constructed from a combination of hybrid and logarithmic terms.
Such a formulation enriches the cosmological dynamics of the model and provides additional
freedom to satisfy the energy conditions, while offering a consistent description of the
evolution of the Universe within the teleparallel framework.

\subsection{Energy Conditions in Hybrid  Teleparallel Gravity}

From the expressions of $\rho(t)$ (\ref{eq:rho_t}) and $p(t)$ (\ref{eq:p_t}), one can factor out the same positive quantity in all energy condition combinations:

\begin{equation}
\mathcal{F}(t)
=
\frac{ e^{\gamma \alpha / \sin^2(kt)} }{\kappa^2}
\left( \frac{\alpha}{\sin^2(kt)} \right)^{\sigma-2}
\label{eq:F_factor}
\end{equation}

This factor is strictly positive, so the sign of each energy condition depends only on the remaining bracketed expressions.
\clearpage
\subsubsection*{1. Null Energy Condition (NEC)}

\begin{equation}
\begin{aligned}
\rho + p
=
\mathcal{F}(t)
\Bigg[
&-48 \frac{k^4}{m^3} \frac{\cos(kt)}{\sin^6(kt)}
\left(
\sigma(\sigma-1)
+ \frac{2\gamma\sigma\alpha}{\sin^2(kt)}
+ \frac{\gamma^2\alpha^2}{\sin^4(kt)}
\right) \\
&+ 6 \frac{k^2}{m^2 \sin^2(kt)}
\left( \sigma + \frac{\gamma\alpha}{\sin^2(kt)} \right)
\frac{\alpha}{\sin^2(kt)}
\Bigg]
\end{aligned}
\label{eq:NEC_rhop}
\end{equation}

\noindent
 The sign depends on $\cos(kt)$. A possible violation of the NEC occurs, which is necessary for inflation.

 \begin{figure}[h!]
 	\centering
 	
 	\begin{subfigure}[b]{0.45\textwidth}
 		\centering
 		\includegraphics[width=\textwidth]{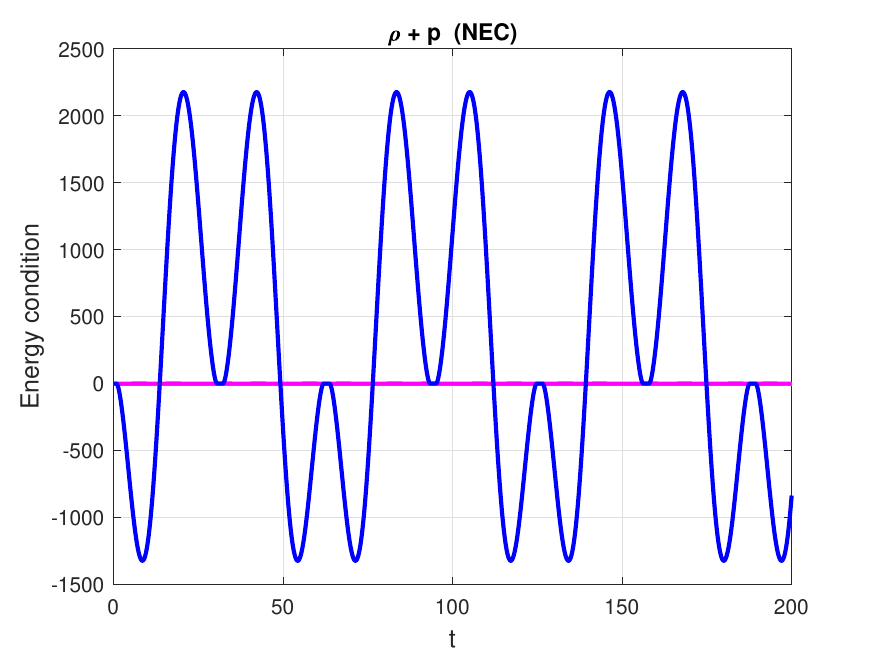}
 		\caption{}

 	\end{subfigure}
 	\hfill
 	\begin{subfigure}[b]{0.45\textwidth}
 		\centering
 		\includegraphics[width=\textwidth]{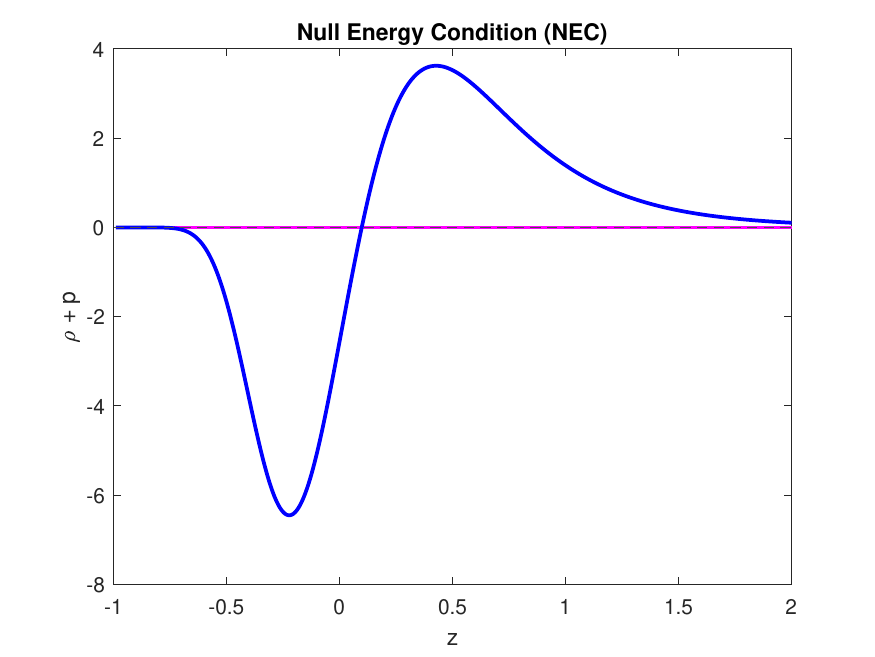}
 		\caption{}
 	\end{subfigure}
 	
 	\caption{Evolution of ECs with $\alpha = 15$, $\beta = -0.5$, and $n = 0$ for $(m,k) = (0.480012, 0.5)$ [green curve], $(m,k) = (0.480003, 0.25)$ [magenta curve], and $(m,k) = (1.15, 0.1)$ [blue curve]: (a) as a function of cosmic time (in Gyr), and (b) as a function of the redshift $z$.}
 \end{figure}

\subsubsection*{2. Dominant Energy Condition (DEC)}

\begin{equation}
\begin{aligned}
\rho - p
=
\mathcal{F}(t)
\Bigg[
&+48 \frac{k^4}{m^3} \frac{\cos(kt)}{\sin^6(kt)}
\left(
\sigma(\sigma-1)
+ \frac{2\gamma\sigma\alpha}{\sin^2(kt)}
+ \frac{\gamma^2\alpha^2}{\sin^4(kt)}
\right) \\
&+ \left(
6 \frac{k^2}{m^2 \sin^2(kt)}
+ 2 \frac{k^2 \cos(kt)}{m \sin^2(kt)}
\right)
\left( \sigma + \frac{\gamma\alpha}{\sin^2(kt)} \right)
\frac{\alpha}{\sin^2(kt)} \\
&+ \frac{\alpha^2}{4 \sin^4(kt)}
\Bigg]
\end{aligned}
\label{eq:DEC_rhominusp}
\end{equation}

\noindent
 Partially satisfied depending on the parameters; compatible with a controlled accelerated phase.
 
 \begin{figure}[h!]
 	\centering
 	
 	\begin{subfigure}[b]{0.45\textwidth}
 		\centering
 		\includegraphics[width=\textwidth]{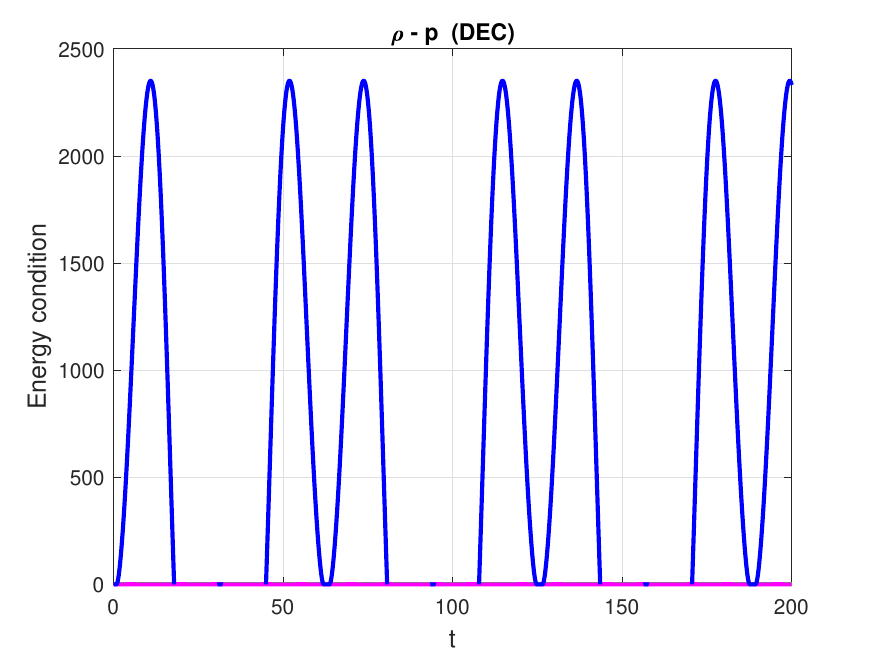}
 		\caption{}
 	\end{subfigure}
 	\hfill
 	\begin{subfigure}[b]{0.45\textwidth}
 		\centering
 		\includegraphics[width=\textwidth]{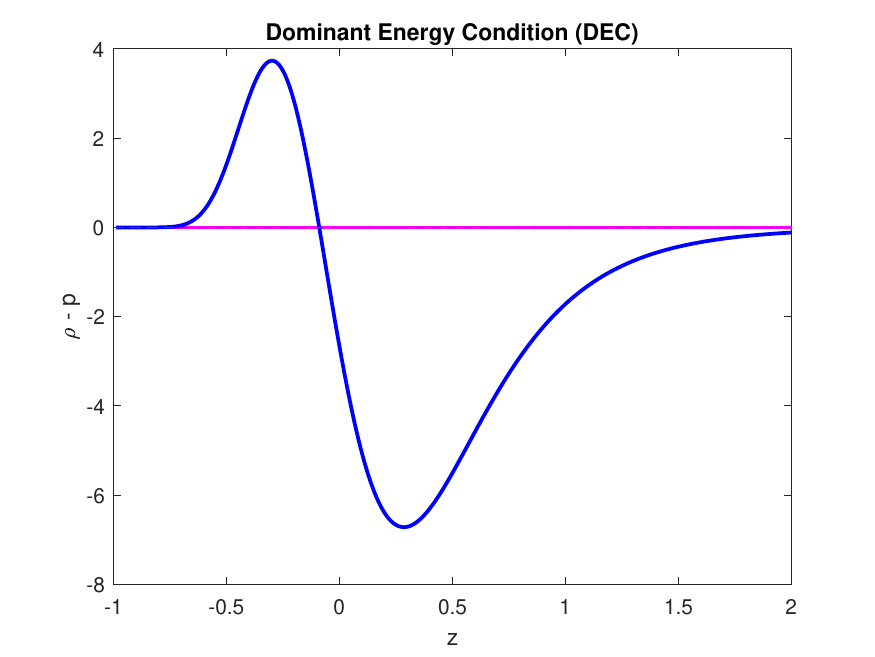}
 		\caption{}
 	\end{subfigure}
 	
 	\caption{Evolution of ECs with $\alpha = 15$, $\beta = -0.5$, and $n = 0$ for $(m,k) = (0.480012, 0.5)$ [green curve], $(m,k) = (0.480003, 0.25)$ [magenta curve], and $(m,k) = (1.15, 0.1)$ [blue curve]: (a) as a function of cosmic time (in Gyr), and (b) as a function of the redshift $z$.}
 \end{figure}

\subsubsection*{3. Strong Energy Condition (SEC)}

\begin{equation}
\begin{aligned}
\rho + 3p
=
\mathcal{F}(t)
\Bigg[
&-144 \frac{k^4}{m^3} \frac{\cos(kt)}{\sin^6(kt)}
\left(
\sigma(\sigma-1)
+ \frac{2\gamma\sigma\alpha}{\sin^2(kt)}
+ \frac{\gamma^2\alpha^2}{\sin^4(kt)}
\right) \\
&- \left(
6 \frac{k^2}{m^2 \sin^2(kt)}
- 6 \frac{k^2 \cos(kt)}{m \sin^2(kt)}
\right)
\left( \sigma + \frac{\gamma\alpha}{\sin^2(kt)} \right)
\frac{\alpha}{\sin^2(kt)} \\
&- \frac{3\alpha^2}{4 \sin^4(kt)}
\Bigg]
\end{aligned}
\label{eq:SEC_rho3p}
\end{equation}
\clearpage

\noindent
 Natural violation of the SEC, a direct signature of dominant negative pressure, necessary for inflation and cosmic acceleration.

 \begin{figure}[h!]
 	\centering
 	
 	\begin{subfigure}[b]{0.45\textwidth}
 		\centering
 		\includegraphics[width=\textwidth]{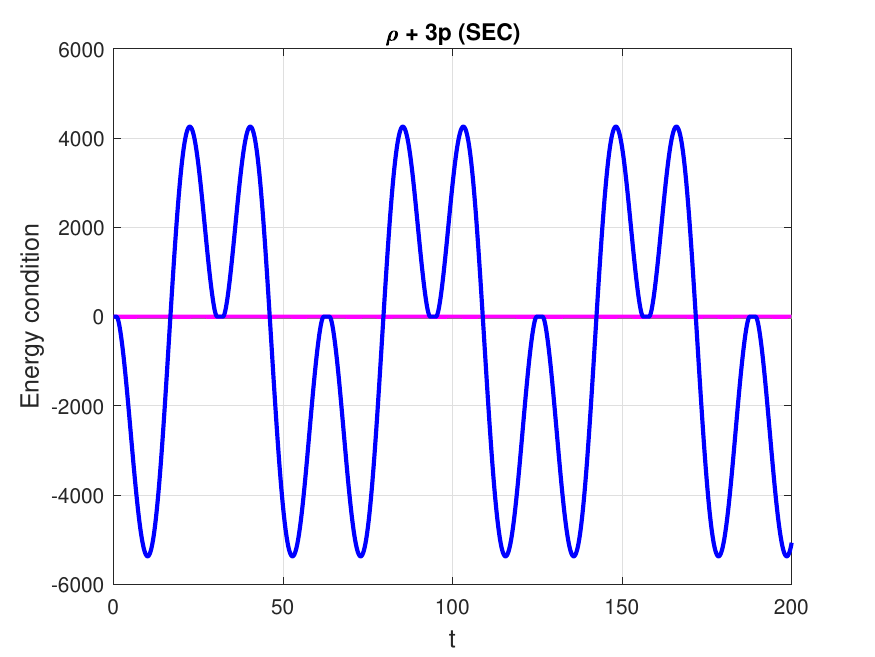}
 		\caption{}
 	\end{subfigure}
 	\hfill
 	\begin{subfigure}[b]{0.45\textwidth}
 		\centering
 		\includegraphics[width=\textwidth]{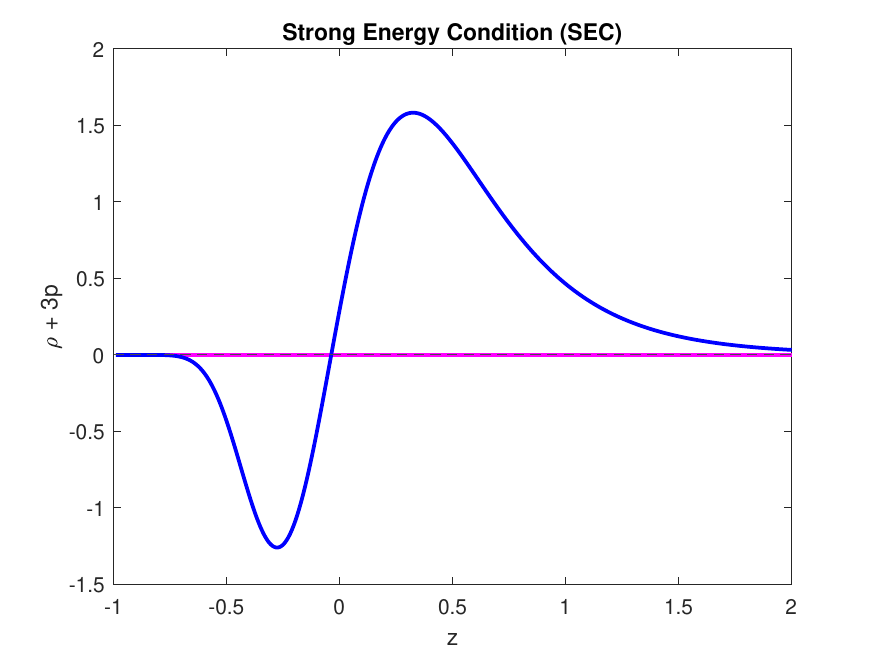}
 		\caption{}
 	\end{subfigure}
 	
 	\caption{Evolution of ECs with $\alpha = 15$, $\beta = -0.5$, and $n = 0$ for $(m,k) = (0.480012, 0.5)$ [green curve], $(m,k) = (0.480003, 0.25)$ [magenta curve], and $(m,k) = (1.15, 0.1)$ [blue curve]: (a) as a function of cosmic time (in Gyr), and (b) as a function of the redshift $z$.}
 \end{figure}
 
 \subsubsection{Graphical Analysis of Energy Conditions in Hybrid Teleparallel Gravity}
 
 In this section, we analyze the behavior of the energy conditions, namely the Null Energy Condition (NEC), Dominant Energy Condition (DEC), and Strong Energy Condition (SEC), in the framework of hybrid teleparallel gravity. The analysis is performed for three different sets of model parameters:
 \begin{equation}
 (m, k) = (0.480012, 0.5) \quad \text{[green curve]}, \quad
 (0.480003, 0.25) \quad \text{[magenta curve]}, \quad
 (1.15, 0.1) \quad \text{[blue curve]}.
 \end{equation}
 
 \paragraph{Null Energy Condition (NEC): $\rho + p \geq 0$}
 
 The evolution of the NEC is depicted in Fig.5 (a) and Fig.5 (b) as a function of cosmic time $t$ and redshift $z$, respectively. As a function of time, the NEC exhibits a clear oscillatory behavior, alternating between positive and negative values. This indicates successive phases where the condition is satisfied and violated. The violation of the NEC ($\rho + p < 0$) suggests the presence of a phantom-like component in the cosmic fluid.
 
 In terms of redshift, the NEC is violated in the recent past (negative $z$ region) and becomes positive at later times, indicating a transition from a phantom to a non-phantom regime. It is important to note that the curves corresponding to the parameter sets $(0.480012, 0.5)$ and $(0.480003, 0.25)$ are almost indistinguishable in several regions, showing a degeneracy in the model parameters. The case $(1.15, 0.1)$ presents a slightly different but qualitatively similar behavior.
 
 \paragraph{Dominant Energy Condition (DEC): $\rho - p \geq 0$}
 
 The DEC is illustrated in Fig.6 (a) and Fig.6 (b). As a function of cosmic time, the DEC remains strictly positive throughout the evolution, ensuring that the energy flow respects causality. This indicates that the model is physically stable in this respect.
 
 However, in terms of redshift, a temporary violation of the DEC is observed in certain intervals, followed by a recovery to positive values at higher redshift. This suggests that causal behavior may be briefly affected during specific cosmological phases.
 
 As in the NEC case, the curves for the first two parameter sets are largely superimposed, confirming a weak dependence of the model on these parameters within this range. The third parameter set $(1.15, 0.1)$ shows more pronounced variations.
 
 \paragraph{Strong Energy Condition (SEC): $\rho + 3p \geq 0$}
 
 The SEC behavior is presented in Fig.7 (a) and Fig.7 (b). The temporal evolution shows strong oscillations with frequent violations of the SEC. Such violations are typically associated with an accelerated expansion of the Universe.
 
 In the redshift representation, the SEC is violated near the present epoch and satisfied in other phases, indicating transitions between accelerated and decelerated expansion regimes.
 
 Again, the curves corresponding to $(0.480012, 0.5)$ and $(0.480003, 0.25)$ overlap significantly, reinforcing the presence of parameter degeneracy. The $(1.15, 0.1)$ case displays higher amplitude oscillations and remains more distinguishable.
 
 %\paragraph{Discussion}
 
 Overall, the hybrid teleparallel gravity model exhibits an oscillatory behavior of the energy conditions, reflecting a rich cosmological dynamics with repeated transitions between phantom and non-phantom phases. The NEC and SEC are violated during certain epochs, consistent with the presence of accelerated expansion and phantom-like behavior.
 
 The DEC is mostly satisfied, ensuring the causal stability of the model, although minor violations may occur in limited intervals. Importantly, the near coincidence of the curves for the first two parameter sets indicates that the model is relatively insensitive to small variations in $m$ and $k$ within this domain.
 
 These results demonstrate that hybrid teleparallel gravity can successfully describe a dynamically evolving Universe with phases of acceleration, while maintaining overall physical consistency.

\subsection{Energy Conditions in Logarithmic  Teleparallel Gravity}

In order to analyze the physical viability of the teleparallel model, we investigate the energy conditions using the effective energy density (\ref{eq:rho_time})  and pressure (\ref{eq:p_time}). For convenience, we define
\begin{equation}
\mathcal{L}(t)
=
\ln\!\left(
\frac{6|b|k^2}{m^2\sin^2(kt)}
\right).
\end{equation}

\subsection*{1. Null Energy Condition (NEC)}

The null energy condition is defined as
\begin{equation}
\rho + p \ge 0.
\end{equation}

Using the above expressions, we obtain
\begin{align}
\rho + p
&=
\frac{D}{4\kappa^2}
\left[
\mathcal{L}-4+4+4m\cos(kt)-\mathcal{L}
\right] \nonumber\\
&=
\frac{Dm}{\kappa^2}\cos(kt).
\end{align}

Thus, the NEC requires
\begin{equation}
\boxed{
	\frac{Dm}{\kappa^2}\cos(kt)\ge0
}
\end{equation}
which is satisfied for $\cos(kt)\ge0$ when $D>0$ and $m>0$.
Hence, the NEC holds in specific temporal intervals and is periodically violated.

\begin{figure}[h!]
	\centering
	
	\begin{subfigure}[b]{0.45\textwidth}
		\centering
		\includegraphics[width=\textwidth]{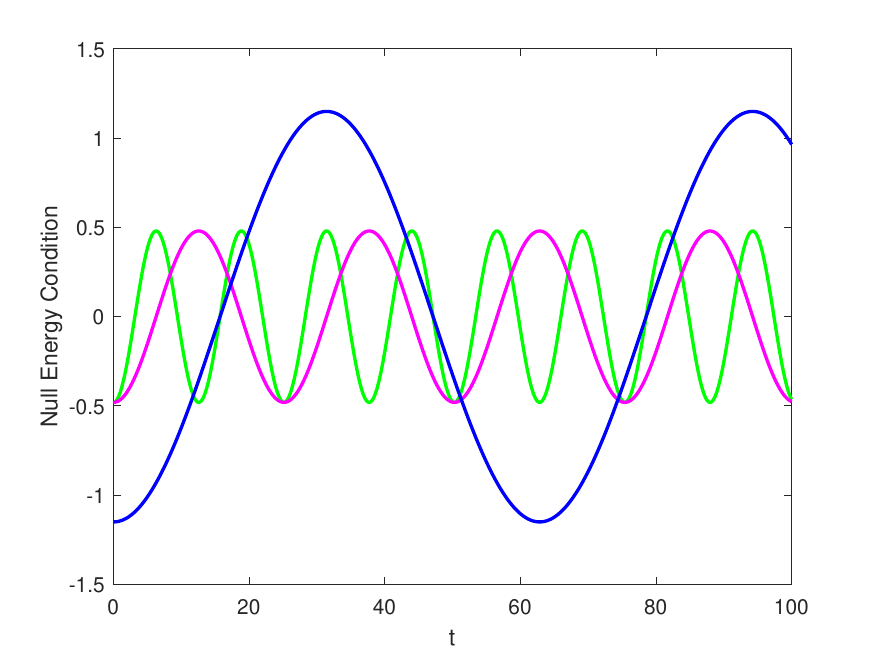}
		\caption{}
	\end{subfigure}
	\hfill
	\begin{subfigure}[b]{0.45\textwidth}
		\centering
		\includegraphics[width=\textwidth]{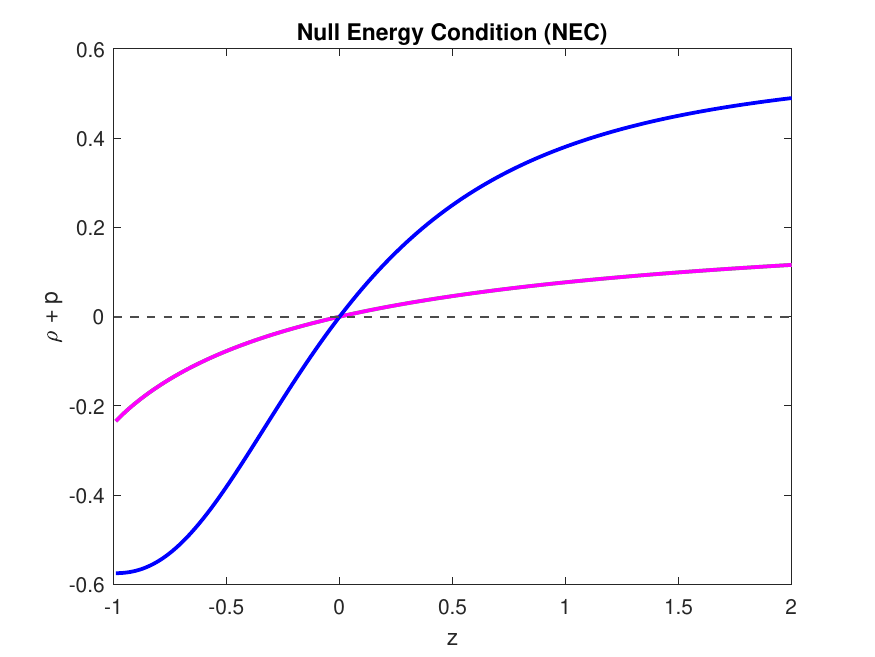}
		\caption{}
	\end{subfigure}
	
	\caption{Evolution of ECs with $\alpha = 15$, $\beta = -0.5$, and $n = 0$ for $(m,k) = (0.480012, 0.5)$ [green curve], $(m,k) = (0.480003, 0.25)$ [magenta curve], and $(m,k) = (1.15, 0.1)$ [blue curve]: (a) as a function of cosmic time (in Gyr), and (b) as a function of the redshift $z$.}
\end{figure}

\subsection*{2. Weak Energy Condition (WEC)}
%The weak energy condition

A straightforward calculation gives
\begin{equation}
\boxed{
	\rho - p
	=
	\frac{D}{4\kappa^2}
	\left[
	2\ln\!\left(
	\frac{6|b|k^2}{m^2\sin^2(kt)}
	\right)
	-8
	-4m\cos(kt)
	\right]
}
\label{eq:rho_minus_p}
\end{equation}

The weak energy condition is given by
\begin{equation}
\rho \ge 0,
\qquad
\rho+p \ge 0.
\end{equation}

The second inequality coincides with the NEC. The first one yields
\begin{equation}
\mathcal{L}(t)-4 \ge 0,
\end{equation}
or equivalently
\begin{equation}
\boxed{
	\ln\!\left(
	\frac{6|b|k^2}{m^2\sin^2(kt)}
	\right)\ge4
}
\end{equation}

which leads to
\begin{equation}
\sin^2(kt)
\le
\frac{6|b|k^2}{m^2 e^4}.
\end{equation}

Therefore, the WEC is satisfied in restricted regions of cosmic time,
implying a positive effective energy density during specific oscillatory phases.

\begin{figure}[h!]
	\centering
	
	\begin{subfigure}[b]{0.45\textwidth}
		\centering
		\includegraphics[width=\textwidth]{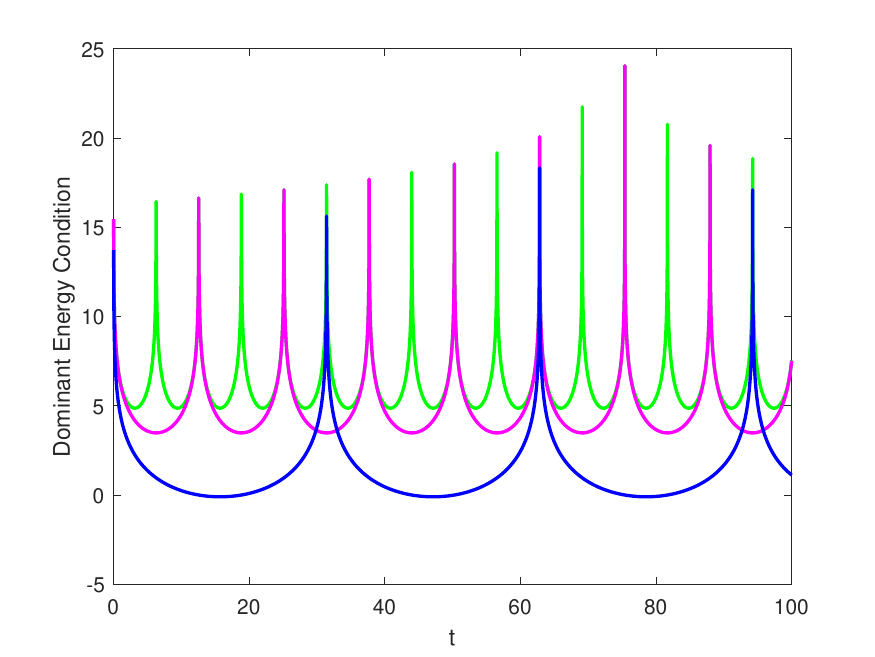}
		\caption{}
	\end{subfigure}
	\hfill
	\begin{subfigure}[b]{0.45\textwidth}
		\centering
		\includegraphics[width=\textwidth]{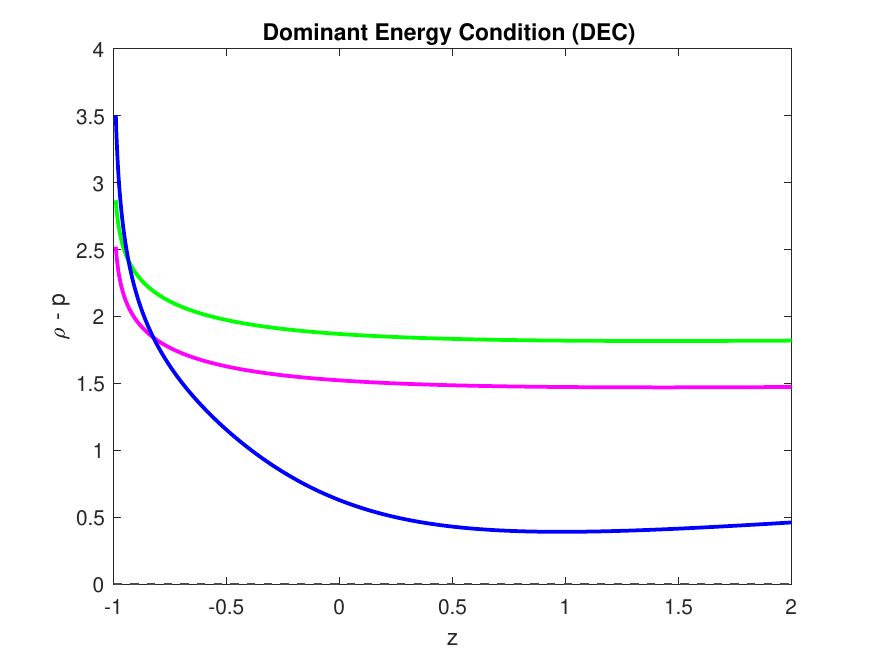}
		\caption{}
	\end{subfigure}
	
	\caption{Evolution of ECs with $\alpha = 15$, $\beta = -0.5$, and $n = 0$ for $(m,k) = (0.480012, 0.5)$ [green curve], $(m,k) = (0.480003, 0.25)$ [magenta curve], and $(m,k) = (1.15, 0.1)$ [blue curve]: (a) as a function of cosmic time (in Gyr), and (b) as a function of the redshift $z$.}
\end{figure}

\subsection*{3. Strong Energy Condition (SEC)}

The strong energy condition is defined as
\begin{equation}
\rho + 3p \ge 0.
\end{equation}

A straightforward calculation gives
\begin{align}
\rho+3p
&=
\frac{D}{4\kappa^2}
\left[
\mathcal{L}-4
+3\left(4+4m\cos(kt)-\mathcal{L}\right)
\right] \nonumber\\
&=
\frac{D}{4\kappa^2}
\left[
8+12m\cos(kt)-2\mathcal{L}(t)
\right].
\end{align}

Thus, the SEC reads
\begin{equation}
\boxed{
	8+12m\cos(kt)
	\ge
	2\ln\!\left(
	\frac{6|b|k^2}{m^2\sin^2(kt)}
	\right)
}
\end{equation}
\clearpage
Due to the logarithmic divergence near $\sin(kt)\rightarrow 0$, the SEC
is violated over wide regions of cosmic evolution.
This violation naturally leads to an accelerated expansion phase
within the teleparallel framework.

\begin{figure}[h!]
	\centering
	
	\begin{subfigure}[b]{0.45\textwidth}
		\centering
		\includegraphics[width=\textwidth]{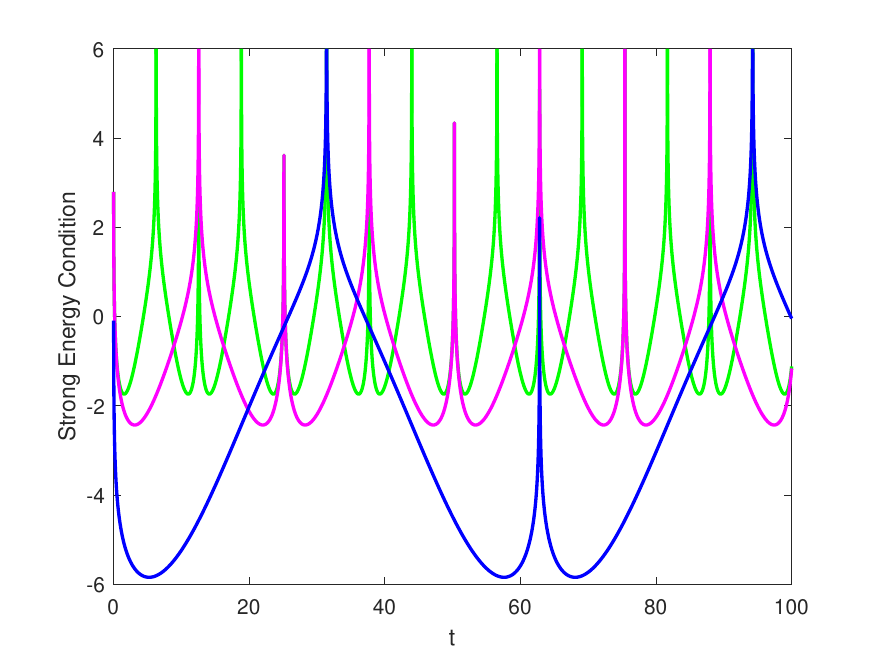}
		\caption{}
	\end{subfigure}
	\hfill
	\begin{subfigure}[b]{0.45\textwidth}
		\centering
		\includegraphics[width=\textwidth]{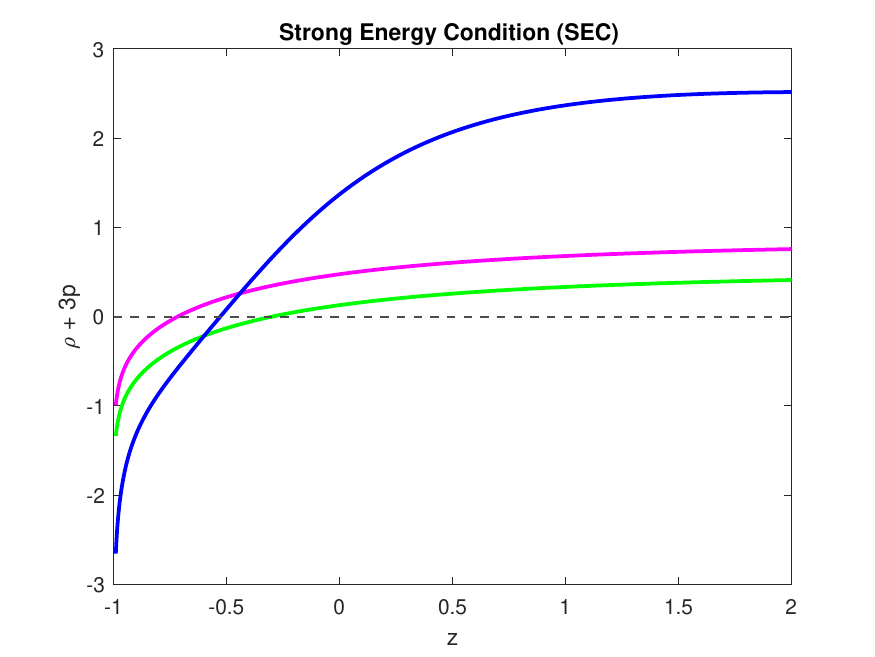}
		\caption{}
	\end{subfigure}
	
	\caption{Evolution of ECs with $\alpha = 15$, $\beta = -0.5$, and $n = 0$ for $(m,k) = (0.480012, 0.5)$ [green curve], $(m,k) = (0.480003, 0.25)$ [magenta curve], and $(m,k) = (1.15, 0.1)$ [blue curve]: (a) as a function of cosmic time (in Gyr), and (b) as a function of the redshift $z$.}
\end{figure}

\subsubsection{Graphical Analysis of Energy Conditions in Logarithmic Teleparallel Gravity}

In this section, we perform a graphical analysis of the energy conditions—namely the Null Energy Condition (NEC), Dominant Energy Condition (DEC), and Strong Energy Condition (SEC)—within the framework of logarithmic teleparallel gravity. The behavior of these conditions is examined through their graphical evolution for three distinct sets of parameters:
\begin{equation}
(m, k) = (0.480012, 0.5) \ \text{[green curve]}, \quad
(0.480003, 0.25) \ \text{[magenta curve]}, \quad
(1.15, 0.1) \ \text{[blue curve]}.
\end{equation}

\paragraph{Null Energy Condition (NEC): $\rho + p \geq 0$}

The evolution of the NEC is presented as a function of cosmic time $t$ and redshift $z$. In the time domain, the NEC exhibits an oscillatory behavior for all parameter sets. The green and magenta curves show small-amplitude oscillations around zero, indicating alternating phases of satisfaction and violation of the NEC. In contrast, the blue curve displays larger amplitude oscillations, with more pronounced violations.

In terms of redshift, the NEC is initially violated in the past ($z < 0$), particularly for the blue curve, and then becomes positive as the Universe evolves toward higher redshift values. This transition reflects a change from a phantom regime to a non-phantom phase. The green and magenta curves are nearly superimposed, indicating that the model is weakly sensitive to small variations in $(m, k)$ within this range.

\paragraph{Dominant Energy Condition (DEC): $\rho - p \geq 0$}

The DEC behavior shows distinct features. As a function of time, all curves remain strictly positive despite the presence of sharp peaks, indicating that the dominant energy condition is always satisfied. These peaks suggest rapid variations in the effective energy density and pressure but do not lead to violations.

In the redshift representation, the DEC remains positive for all parameter sets, confirming that the model preserves causal energy propagation. The blue curve starts from a higher value and decreases smoothly, while the green and magenta curves exhibit a more gradual evolution. Again, the green and magenta curves are closely aligned, reinforcing the degeneracy between these parameter choices.

\paragraph{Strong Energy Condition (SEC): $\rho + 3p \geq 0$}

The SEC displays a more complex behavior. In the time evolution, strong oscillations are observed, with both positive and negative values. The frequent violations of the SEC indicate the presence of accelerated expansion phases. The blue curve shows deeper negative regions, implying a stronger deviation from standard matter behavior.

As a function of redshift, the SEC is initially violated (negative values) and then becomes positive as $z$ increases. This behavior indicates a transition from an accelerated expansion phase to a decelerated regime. The green and magenta curves are again nearly indistinguishable, while the blue curve shows a steeper evolution.

\paragraph{Discussion}

Overall, the logarithmic teleparallel gravity model exhibits a rich dynamical structure characterized by oscillatory behavior in the energy conditions. The NEC and SEC are violated during certain epochs, which is consistent with the presence of phantom-like behavior and accelerated expansion. 

The DEC remains satisfied throughout the evolution, ensuring the physical viability and causal consistency of the model. Importantly, the near coincidence of the green and magenta curves indicates a degeneracy in the parameter space, suggesting that small variations in $(m, k)$ do not significantly affect the cosmological dynamics.

The blue curve $(1.15, 0.1)$, however, shows more pronounced deviations, highlighting the sensitivity of the model to larger changes in the parameters. These results demonstrate that logarithmic teleparallel gravity can successfully describe different cosmological phases, including transitions between phantom and non-phantom regimes.

\section{Conclusion}

In this work, we have investigated a cosmological model within the framework of modified teleparallel gravity by considering two specific functional forms of $ f(T)$ , namely a hybrid model of the form $ f(T)=e^{\gamma T}T^{\sigma} $ and a logarithmic model, in the context of a periodic cosmic evolution driven by an oscillating deceleration parameter.

By adopting the parametrization $ q(t)=m\cos(kt)-1 $, we have shown that the model is capable of reproducing a cyclic cosmic dynamics characterized by successive transitions between decelerating and accelerating phases. In particular, for values consistent with current observations, such as $ m \simeq 0.48 $ and $ H_0 = 69.2 \,\text{km s}^{-1}\text{Mpc}^{-1} $, the model successfully describes a late-time accelerating phase with $ q_0 \approx -0.52 $. For $ m \geq 1 $, a strongly oscillatory behavior emerges, including phases of super-accelerated expansion ($ q <-1 $).

In the case of the hybrid model, for the parameter values $ \gamma = 0.1 $ and $ \sigma = -0.5 $, the analysis of cosmological quantities shows that the energy density $ \rho(t) $ remains globally positive throughout the evolution, while the pressure $ p(t) $ exhibits an oscillatory behavior with recurrent sign transitions. The equation of state parameter $ w(t) $ evolves dynamically and crosses both the quintessence ($ -1<w<-1/3 $) and phantom ($ w<-1 $) regimes, confirming the ability of the model to describe different phases of dark energy.

Regarding the logarithmic model, we have shown that logarithmic corrections play a crucial role in stabilizing the cosmic dynamics. In contrast to the hybrid model, the logarithmic contribution tends to regularize divergences occurring near the points where $\sin(kt) \to 0 $, leading to a smoother evolution of the cosmological parameters. In particular, the equation of state parameter exhibits milder oscillations and remains longer in the quintessence regime, thereby reducing extreme phantom phases.

The graphical analysis of the energy conditions (NEC, DEC, and SEC) indicates that their validity strongly depends on the choice of parameters $ (m,k,\gamma,\sigma) $. In several configurations, the violation of the SEC is observed, which is consistent with an accelerated expansion phase. Meanwhile, the NEC and DEC are partially satisfied, supporting the physical viability of the model over certain cosmological epochs.

Overall, our results demonstrate that the combined introduction of a periodic deceleration parameter and hybrid and logarithmic corrections in teleparallel gravity provides a rich and flexible cosmological framework compatible with current observations. The model thus offers an interesting alternative to the standard $\Lambda$CDM scenario, while providing a unified description of the different phases of the cosmic expansion.

\end{document}